\documentclass[journal,draftcls,onecolumn,12pt,twoside]{IEEEtranTCOM}

\usepackage{amsmath}
\usepackage{amsthm}
\usepackage{amssymb}
\usepackage{color}
\usepackage[pdftex]{graphicx}
\usepackage{graphicx}
\usepackage{epstopdf}

\newtheorem{thm}{Theorem}

\newtheorem{rem}{Remark}

\def\p0{{\pmb 0}}

\long\def\comment#1{}

\newfont{\bbb}{msbm10 scaled 700}

\newfont{\bbc}{msbm10 scaled 1100}

\newcommand{\EE}{\mbox{\bbc E}}

\newcommand{\av}{{\pmb a}}
\newcommand{\bvx}{{\pmb b}}

\newcommand{\ev}{{\pmb e}}

\newcommand{\gv}{{\pmb g}}
\newcommand{\hv}{{\pmb h}}

\newcommand{\pvx}{{\bf p}}
\newcommand{\qv}{{\pmb q}}
\newcommand{\rv}{{\pmb r}}
\newcommand{\sv}{{\pmb s}}

\newcommand{\uv}{{\pmb u}}
\newcommand{\vv}{{\pmb v}}

\newcommand{\xv}{{\pmb x}}
\newcommand{\yv}{{\pmb y}}
\newcommand{\zv}{{\pmb z}}
\newcommand{\zerov}{{\pmb 0}}
\newcommand{\onev}{{\pmb 1}}

\newcommand{\Am}{{\pmb A}}
\newcommand{\Bm}{{\pmb B}}
\newcommand{\Cm}{{\pmb C}}
\newcommand{\Dm}{{\pmb D}}

\newcommand{\Gm}{{\pmb G}}

\newcommand{\Id}{{\pmb I}}

\newcommand{\Km}{{\pmb K}}

\newcommand{\Pm}{{\pmb P}}

\newcommand{\Tm}{{\pmb T}}

\newcommand{\Vm}{{\pmb V}}

\newcommand{\Xm}{{\pmb X}}

\newcommand{\Zm}{{\pmb Z}}

\newcommand{\Cc}{{\cal C}}

\newcommand{\Nc}{{\cal N}}

\newcommand{\betav}{\hbox{\boldmath$\beta$}}

\newcommand{\phiv}{\hbox{\boldmath$\phi$}}

\newcommand{\Lambdam}{\hbox{\boldmath$\Lambda$}}

\newcommand{\Phim}{\hbox{\boldmath$\Phi$}}

\newcommand{\diag}{{\hbox{diag}}}

\newcommand{\herm}{{\sf H}}

\begin{document}

\title{Uplink Interference Reduction in Large Scale Antenna Systems }


\author{\IEEEauthorblockN{Ansuman Adhikary\IEEEauthorrefmark{1},
Alexei Ashikhmin\IEEEauthorrefmark{2} and Thomas L. Marzetta\IEEEauthorrefmark{2}}
\thanks{\IEEEauthorrefmark{1} A. Adhikary is with Ericsson Inc, 200 Holger Way, San Jose, CA 95134.}
\thanks{\IEEEauthorrefmark{2} A. Ashikhmin and T. L. Marzetta are with Bell Laboratories, Nokia, 600 Mountain Ave, Murray Hill, NJ 07974.}}

\maketitle

\vspace{-2.5cm}
\begin{abstract}
A massive MIMO system entails a large number (tens or hundreds) of base station
antennas serving a much smaller number of terminals. These systems demonstrate large
gains in spectral and energy efficiency compared with
conventional MIMO technology. As the number of antennas grows, the
performance of a massive MIMO system  gets limited by the interference caused by pilot contamination \cite{marzetta2010noncooperative}.
In \cite{ashikhmin2012pilot}, \cite{ashikhminPart1} A.~Ashikhmin and T.~Marzetta proposed  (under the name of Pilot Contamination Precoding)
Large Scale Fading Precoding (LSFP) and Decoding (LSFD) based on limited cooperation between base stations.
They showed that Zero-Forcing LSFP and LSFD eliminate pilot
contamination entirely and lead to an infinite throughput as the number of antennas grows.

 In this paper, we focus on the uplink and show
that even in the case of a finite number of base station antennas,
LSFD yields a very large performance gain. In particular, one of our algorithms gives a more than 140
fold increase in the $5$\% outage data transmission rate! We
show that the performance can be improved further by optimizing the
transmission powers of the users. Finally, we present decentralized LSFD that
requires limited cooperation only between neighboring cells.
\end{abstract}


\section{Introduction} \label{sec:intro}
In recent years, massive MIMO systems have become
quite promising in terms of meeting the increasing demand to enable
high data rates in cellular systems, see  \cite{massiveMIMOreview} and references within.
In a massive MIMO system, the base station (BS)
is equipped with a very large number of antennas that significantly
exceeds the number of users. It was shown in
\cite{marzetta2010noncooperative} that when the number of antennas
tends to infinity, the main limiting factor in performance is pilot
contamination, which arises due to the fact that the users in different cells
unavoidably use nonorthogonal pilot signals during estimation of the channel.A number of works in the literature have been devoted to the use of efficient schemes in order to mitigate the pilot contamination effect, see for example \cite{pilotcontref1}-\cite{ashikhminPart1} and references therein. The works in \cite{pilotcontref1}-\cite{pilotcontref3} assume the channel to be low rank due to the presence of a smaller number of multipath components when compared with the number of antennas. Due to this reason, the channel matrices of the users span a low rank subspace. Exploting this idea, efficient precoders have been designed to make the resultant channel between the users orthogonal thereby mitigating the pilot contamination effect. The works
\cite{ashikhmin2012pilot}, \cite{ashikhminPart1}, however, do not make any assumption on the low rank property of the channel and alleviate the pilot contamination problem
through the use of Large Scale Fading Precoding/Decoding schemes, referred
to as LSFP and LSFD respectively. LSFP and LSFD assume two stage precoding/decoding. In particular, in LSFD, at the first stage,  each base
station equipped with $M$ antennas conducts $M$-dimensional MIMO decoding of received signals in order to get estimates of transmitted uplink signals. For instance, base station can use $M$-dimensional matched filtering,
zero-forcing, or MMSE MIMO decoders. This stage is conducted completely locally and does not require any cooperation between base stations.
At the second stage, each base station forwards the obtained uplink signal estimates  to a network controller.
The network controller uses this information for conducting an $L$-dimensional large scale
fading coefficients decoding, where $L$ is the number of cells in the network. This decoding involves only large scale fading coefficients.

It is important to note that large-scale fading coefficients
do not depend on antenna and OFDM frequency subcarrier index. Thus, between any base station and user,
   there is only one such coefficient. Therefore LSFD requires only small bandwidth on backhaul link between base stations and the network controller, and
this bandwidth does not grow with $M$. Further,   in
the radius of 10 wavelengths the large-scale
fading coefficients are approximately constant (see \cite{huang} and references there), while small-scale fading coefficients
significantly change as soon as a user moves by a quarter of the wavelength. Thus,
large-scale fading coefficients change about 40 times slower and, for this reason, LSFD is robust to
user mobility.

It is shown in \cite{ashikhmin2012pilot} that when the number of antennas grows to infinity and the number of cells $L$ stays
   constant, Zero-Forcing LSFD (ZF LSFD) allows one to completely cancel interference and provides each user with SINR that grows linearly
    with the number of antennas. In real life scenarios, however, when the number of antennas $M$ is finite, other sources of
    interferences, beyond the one caused by pilot contamination, are significant. As a result, ZF LSFD begins providing
     performance gain only at very large number of antennas, like $M>10^5$. In contrast, at a smaller number of antennas, ZF LSFD
      results in system performance degradation compared with the case when no cooperation between base stations is used.

    A natural question therefore is to ask
    whether one can design LSFD so that LSFD would improve system performance for relatively small values of $M$, or LSFD is only a theoretical tool useful
    for analysis of asymptotic regimes.  In this work, we design LSFDs that take into account {\em all} sources of interference
     and show that such LSFDs provide performance gain in the case of any finite $M$ (we are specifically interested in scenarios when
     $M$ is around $100$).

     As performance criteria, we use the {\em minimum rate} among all users and the {\em $5\%$-outage rate}, which is the smallest rate among $95\%$ of the best users.
 In future generations of wireless systems, all or almost all users will have to be served with large rates.
 Therefore, we believe that these criteria are more meaningful than the often used sum throughput. For optimizing the above criteria,
 we consider max-min optimization problems. Though max-min optimization is strictly speaking not optimal for $5\%$-outage rate criterion, it gives
 very good results, and therefore can be considered  as an engineering tool for optimizing the $5\%$-outage rate.

 {\em Notation : } We use boldface capital and small letters for matrices and vectors respectively. $\Xm^T$, $\Xm^\herm$ and $\Xm^{-1}$ denote the transpose, Hermitian transpose and inverse of $\Xm$, $||\xv||$ denotes the vector 2-norm of $\xv$, and the identity matrix is denoted by $\Id$.


\section{System Model} \label{sec:sys-model}

We consider a multicell system comprised of $L$ cells with each cell having a BS equipped with
$M$ antennas and serving $K$ single antenna users with random locations in the corresponding cell.
 We assume that the network uses frequency reuse factor 1
and consider a flat fading channel
model for each OFDM subcarrier. In what follows, we omit the
subcarrier index and focus on a single subcarrier. For a given subcarrier,  the $M \times 1$
{\em channel vector} between the $k^{\rm th}$ user in the $l^{\rm th}$ cell to
the BS in the $j^{\rm th}$ cell is denoted by
 \begin{equation}
 \gv_{jkl} = \sqrt{\beta_{jkl}} \hv_{jkl}
 \end{equation}
 where $\beta_{jkl}$ denotes the {\em large scale  fading coefficient} that depends on the user location and the propagation environment between the user and the BS, and $\hv_{jkl}=(h_{jkl1},\ldots h_{jklM})^T$
 denotes the {\em small scale fading vectors} whose entries  $h_{jklm},m=1,M$, are  {\em small scale fading coefficients}. We assume that
  $\hv_{jkl} \sim \Cc\Nc(\zerov,\Id_M)$. The coefficients  $\beta_{jkl}$ are modeled, according to \cite{3gppscm}, as
 \begin{equation} \label{eqn:pathloss}
 10 \log_{10} (\beta_{jkl}) = -127.8 - 35 \log_{10}(d_{jkl}) + X_{jkl}
 \end{equation}
 where $d_{jkl}$ denotes the distance (in km) between the user and base station and $X_{jkl} \sim \Cc \Nc (0, \sigma_{\rm shad}^2)$, where the variance
 $\sigma_{\rm shad}^2$ represents the shadowing.

We assume a time block fading model. Thus, small scale fading vectors
$\hv_{jkl}$ stay constant during the {\em coherence interval}. It is convenient to measure the length $T$ of the coherence interval in terms of
 the number of OFDM symbols that can be transmitted within that interval.
 Similarly  large scale fading coefficients $\beta_{jkl}$ stay constant during {\em large scale coherence interval} of $T_\beta$ OFDM
 symbols.  A usual assumption is that $T_\beta$ is about 40 times larger than
 $T$. The vectors $\hv_{jkl}$ and coefficients $\beta_{jkl}$ are assumed to be independent in different
coherence intervals and large scale coherence intervals respectively.

Finally, we assume reciprocity between uplink
and downlink channels, i.e., $\beta_{jkl}$ and $\hv_{jkl}$ are the same for these channels.

 It is important to note that small scale fading coefficients $h_{jklm}$ depend on antenna index and
 on OFDM subcarrier index. If $\Delta$ is the number of OFDM tones in
  the coherence bandwidth and $N$ is the total number of OFDM tones, then
  between a BS and a user, there are
 $MN/\Delta$ small scale fading coefficients and only one large scale fading coefficient.

\section{Time Division Protocol}\label{sec:tdd}
We assume that in all cells, the same set of orthonormal training sequences $\phiv_1, \ldots,\phiv_K\in {\mathbb C}^{1\times K},~\phiv_i \phiv_j^H=\delta_{ij}$, are used.
We assume that
in each cell, the users are enumerated and that the $k^{\rm th}$user uses the training sequence
$\phiv_k$.  We will use the notation $\Phim=\left(\begin{array}{c}\phiv_1\\ \vdots\\\phiv_K\end{array}\right)$.

\begin{rem}
One can also consider a system in which
different sets of pilots are used in different cells. We leave this interesting topic for future work.
\end{rem}

\begin{figure}[!htb]
 \vspace{-6cm}
  \centering
\includegraphics[scale=0.5]{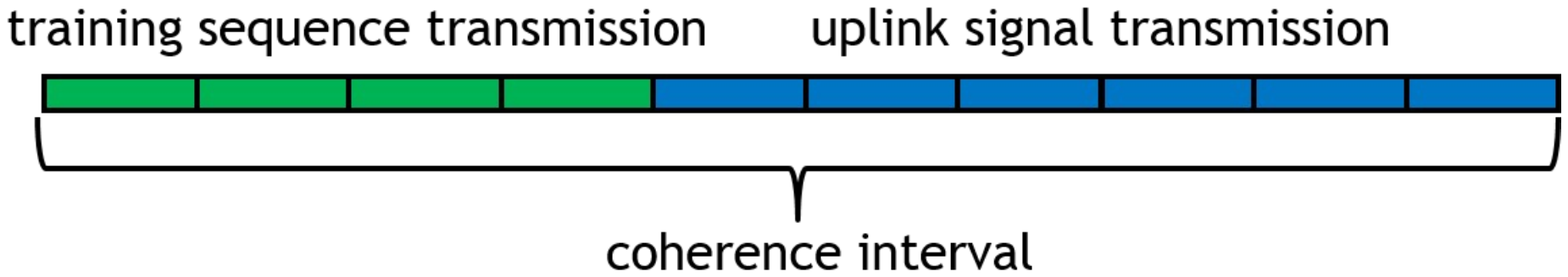}
\vspace{-7cm}
  \caption{TDD protocol with $T=10$}\label{fig:TDDprotocol}
\end{figure}

Let $s_{kl}$ be the {\em uplink signal} transmitted by the $k^{\rm th}$ user in the $l^{\rm th}$ cell.
The TDD communication protocol consists of the following two steps as shown in Fig. \ref{fig:TDDprotocol}.

\noindent{\bf TDD Protocol}
\begin{enumerate}
\item all users synchronously transmit their training sequences $\phiv_k,\;k=1,K$;
\item The $l^{\rm th}$ BS uses the received training sequences to get MMSE estimates $\hat{\gv}_{lkl}$ of the vectors $\gv_{lkl},k=1,\ldots,K$;
\item all users synchronously transmit their uplink signals $s_{kl},\;k=1,K,\;l=1,L$;
\item The $l^{\rm th}$ BS conducts an $M$-dimensional MIMO decoding of the received uplink signals and gets estimates $\tilde{s}_{kl}$ of $s_{kl},k=1,\ldots,K$.
\item BS $l$ transmits the estimates $\tilde{s}_{kl}$ to the network controller via a backhaul link.
 \end{enumerate}
\noindent{\bf The end.}

 At the first step of the TDD protocol, the $l^{\rm th}$ BS receives the signal
 \begin{equation}
 \Tm_l = \sum_{n=1}^L \Gm_{ln} \Pm_n^{\frac{1}{2}} \Phim + \Zm_l
 \end{equation}
 where $\Gm_{ln} = [\gv_{l1n} \gv_{l2n} \ldots \gv_{lKn}]$ is the concatenation of the user channel vectors in the $n^{\rm th}$ cell to the BS in the $l^{\rm th}$ cell,
 $\Pm_n=\diag(p_{1n},\ldots,p_{Kn})$ is the diagonal channel matrix of {\em the training powers} $p_{kn}$ used during the uplink training phase in the $n^{\rm th}$ cell and $\Zm_l$ is AWGN with entries that are i.i.d. $\Cc\Nc(0,1)$ random variables.

 Multiplying $\Tm_l$ by $\Phim^\herm$ and extracting the $k^{\rm th}$ column of $\Tm_l \Phim^\herm$, the $l^{\rm th}$ BS gets
 \begin{equation}
 \rv_{kl} = \sum_{n=1}^L \gv_{lkn} \sqrt{p_{kn}} + \bar{\zv}_l
 \end{equation}
 where $\bar{\zv}_l \sim \Cc\Nc(\zerov,\Id_M)$.

 The MMSE estimate $\hat{\gv}_{lkl}$ of the channel vector $\gv_{lkl}$ is given by
 \begin{eqnarray}
 \hat{\gv}_{lkl} &=& \EE[\gv_{lkl} \rv_{kl}^\herm] \EE[\rv_{kl} \rv_{kl}^\herm]^{-1} \rv_{kl}
 = \frac{\beta_{lkl} \sqrt{p_{kl}}}{1 + \sum_{n=1}^L \beta_{lkn} p_{kn}}
 \rv_{kl}\label{eq:hhat_lkl}
 \end{eqnarray}
 Denote by $\ev_{lkl}$ the estimation error. Then,
 $\gv_{lkl} = \hat{\gv}_{lkl} + \ev_{lkl}$.
 It is well known that
$\ev_{lkl}$ is independent and uncorrelated with $\hat{\gv}_{lkl}$ and
\begin{equation}
 \hat{\gv}_{lkl} \sim \Cc\Nc\left(\zerov,\frac{\beta_{lkl}^2 p_{kl}}{1 + \sum_{n=1}^L
\beta_{lkn} p_{kn}} \Id_M\right),~
 \ev_{lkl} \sim \Cc\Nc \left(\zerov,
\left(\beta_{lkl} - \frac{\beta_{lkl}^2 p_{kl}}{1 + \sum_{n=1}^L
\beta_{lkn} p_{kn}} \right)\Id_M\right) \label{eq:C_hv}
\end{equation}
Invoking the MMSE decomposition, we can write
 $\gv_{lkm} = \hat{\gv}_{lkm} + \ev_{lkm}$,
 where, using (\ref{eq:hhat_lkl}), we have
 \begin{align} \label{eqn:mmse-1}
 \hat{\gv}_{lkm} &= \EE[\gv_{lkm} \rv_{kl}^\herm] \EE[\rv_{kl} \rv_{kl}^\herm]^{-1} \rv_{kl}
 = \frac{\beta_{lkm} \sqrt{p_{km}}}{\beta_{lkl} \sqrt{p_{kl}}} \frac{\beta_{lkl} \sqrt{p_{kl}}}{1 + \sum_{n=1}^L \beta_{lkn} p_{kn}} \rv_{kl}
 = \frac{\beta_{lkm} \sqrt{p_{km}}}{\beta_{lkl} \sqrt{p_{kl}}}
 \hat{\gv}_{lkl},
 \end{align}
 and
 $$
 \ev_{lkm} \sim \Cc\Nc \left(\zerov, \left(\beta_{lkm} - \frac{\beta_{lkm}^2 p_{km}}{1 + \sum_{n=1}^L \beta_{lkn} p_{kn}} \right)\Id_M\right)
 $$
 According to the TDD protocol, the signal received by the $l^{\rm th}$ BS at the third step
  is
 \begin{equation}\label{eq:yv_l}
 \yv_l = \sum_{n=1}^L \sum_{m=1}^K \gv_{lmn} \sqrt{q_{mn}} s_{mn} + \zv_l
 \end{equation}
 where $q_{mn}$ is {\em the transmit power} of the $m^{\rm th}$ user in the $n^{\rm th}$ cell and $s_{mn}$ is its data symbol.

We can
use several possible $M$-dimensional decoding algorithms
for getting estimates of $s_{kl}$ from $\yv_l$. In particular, we
can use matched filtering, zero-forcing or MMSE decoding.

The matched filtering operation has the smallest computational complexity among these three decoding algorithms.
In addition, matched filtering does not require any cooperation between BS antennas and thus, significantly
simplifies  base station hardware.
If the $l^{\rm th}$ BS uses matched filtering, then it gets for the $k^{\rm
th}$ user of $l^{\rm th}$ cell the estimate

\begin{align}
 \tilde{s}_{kl} = &\ \hat{\gv}_{lkl}^\herm \yv_l
 = \sum_{n=1}^L \hat{\gv}_{lkl}^\herm \gv_{lkn} \sqrt{q_{kn}} s_{kn} + \sum_{n=1}^L \sum_{m=1,m \neq k}^K \hat{\gv}_{lkl}^\herm \gv_{lmn} \sqrt{q_{mn}} s_{mn}
  + \hat{\gv}_{lkl}^\herm \zv_l \label{eq:s_tilde0}\\
=\ & \underbrace{\EE [\hat{\gv}_{lkl}^\herm \gv_{lkl}] \sqrt{q_{kl}} s_{kl}}_{\rm Useful\ \ Term} +
\underbrace{\sum_{n=1, n\neq l}^L \EE[\hat{\gv}_{lkl}^\herm \gv_{lkn}] \sqrt{q_{kn}} s_{kn}}_{\rm Pilot Contamination Term} \nonumber\\
& + \underbrace{\sum_{n=1}^L \left(\hat{\gv}_{lkl}^\herm \gv_{lkn} -
\EE[\hat{\gv}_{lkl}^\herm \gv_{lkn}] \right) \sqrt{q_{kn}} s_{kn}
+\sum_{n=1}^L \sum_{m=1,m \neq k}^K \hat{\gv}_{lkl}^\herm
\gv_{lmn} \sqrt{q_{mn}} s_{mn} + \hat{\gv}_{lkl}^\herm
\zv_l}_{\rm Interference \ \ + \ \
 Noise\ \ Terms} \label{eq:stilde_kl}
\end{align}
It is not difficult to see  that in (\ref{eq:stilde_kl}),  the power of the useful
term is proportional to $\left| \EE [\hat{\gv}_{lkl}^\herm \gv_{lkl}] \right|^2$
and therefore is proportional to $M^2$. The powers of the pilot
contamination terms are also proportional to $M^2$. At the same time,
the powers of all other terms are proportional only to $M$. These observations, after some additonal analysis, lead to the following result obtained in
\cite{marzetta2010noncooperative} (see also \cite{fernand2012}, \cite{fernand2013}).
\begin{thm}\label{thm:finite_SINR}
Assuming $p_{kl}=q_{kl}$ we have
$
\lim_{M\to\infty}    \mbox{SINR}_{kj} \stackrel{\textrm{a.s.}}{=}
 {q_{kj}\beta_{jkj}^2
 \over
 \sum_{l=1 \atop l\not =i}^L q_{kl}\beta_{jkl}^2
 }.
$
\end{thm}

\section{Large Scale Fading Decoding}\label{sec:analysis-pcp}

Several techniques, such as power allocation algorithms, frequency reuse schemes,  and others,
have been proposed to mitigate the effect of pilot contamination, see \cite{massiveMIMOreview}, \cite{fernand2012}, \cite{fernand2013}. These techniques allow one to mitigate the pilot contamination interference, but neither of them completely eliminates it.
As a result, similar to Theorem \ref{thm:finite_SINR}, the SINRs stay finite
even in the asymptotic regime as $M$ tends to infinity.

 For obtaining a system in which SINRs grow along with $M$, one may try to use a network MIMO scheme (see for example \cite{NetworkMIMO1}, \cite{NetworkMIMO2}, \cite{NetworkMIMO3}). In such a system,
 the $j^{\rm th}$ base station estimates the coefficients
$\beta_{jkl}$ and $h_{jkl}$ for $k=1,K,~l=1,L$, and $m=1,M$, and sends them
to the network controller (or other base stations). This allows all base stations to behave as one super large antenna array.
This approach, however, seems to be infeasible for the following reason.

It can be seen that the number of small scale fading coefficients $h_{jklm}$
 is proportional to $M$. Thus, in the asymptotic regime, as $M$ tends to infinity, the needed backhaul bandwidth grows infinitely, and the network MIMO scheme becomes infeasible.
Even in the case of finite $M$, the needed backhaul bandwidth is tremendously large. For instance, assuming $M=100$, the coherence bandwidth $\Delta=14$ and the number of OFDM tones $N=1400$,  we obtain that
 the $j^{\rm th}$ base station needs to transmit to the network controller
$NM/\Delta\cdot K(L-1)=10000K(L-1)$ small scale fading coefficients. Note also that typically coherence interval is short, i.e., $T$ is small, since the small scale fading coefficients substantially change as soon as a mobile moves a quarter of the wavelength. Thus, those $10000K(L-1)$
coefficients will be sent quite frequently. All of these  make the needed backhaul bandwidth hardly feasible.


A breakthrough  was achieved in
 \cite{ashikhmin2012pilot}, \cite{ashikhminPart1} where it was proposed to organize cooperation
between BSs on the level of large scale fading coefficients in
order to cancel the pilot contamination terms in
(\ref{eq:stilde_kl}). In \cite{ashikhmin2012pilot}, this approach was called Pilot
Contamination Postcoding. Since this approach allows mitigation of all sources of interference, as we show below,
we believe that a more appropriate name for it is
 Large Scale Fading Decoding (LSFD). In Section \ref{sec:results}, we compare LSFD with a Network MIMO scheme in which the pilot contamination is taken into account. Our simulation results show that LSFD has virtually the same performance, while its communication and computation complexities are significantly lower.
 The formal description of LSFD is given below.

\noindent{\bf Large-Scale Fading Decoding}
\begin{enumerate}
\item
 The $l^{\rm th}$ BS estimates
 $\beta_{lkn},k=1,K,~n=1,L$, and  sends them to a controller.
 \item For each $k=1,K$,
 the controller computes the $L\times L$ decoding matrix  $\Am_k
= (\av_{k1} \av_{k2} \ldots
 \av_{kL}),k=1,K$, as functions of $\beta_{lkn},l,n=1,L$.
\item The $l^{\rm th}$ BS computes the MMSE esimates $\hat{\gv}_{lkl}$ according to (\ref{eq:hhat_lkl}).
\item \label{step:M dim decoding} The $l^{\rm th}$ BS receives the vector $\yv_l$ defined in (\ref{eq:yv_l}) and computes signals $\tilde{s}_{kl},k=1,K$, using an $M$-dimensional
decoding, e.g.,
matched filtering (\ref{eq:stilde_kl}), zero-forcing
(\ref{eqn:zf-condn}), or MMSE. It further
sends $\tilde{s}_{kl}$ to the network contoller.
\item The controller forms the vector $\tilde{\sv}_k=[\tilde{s}_{k1},\ldots,\tilde{s}_{kL}]^T$ and computes
 the estimates $\hat{s}_{kl}=\av_{kl}^\herm \tilde{\sv}_k,~k=1,K,~l=1,L$.
 \end{enumerate}
\noindent{\bf The end.}

\begin{figure}[!htb]
  \centering
\hspace{-0cm}\includegraphics[scale=0.35]{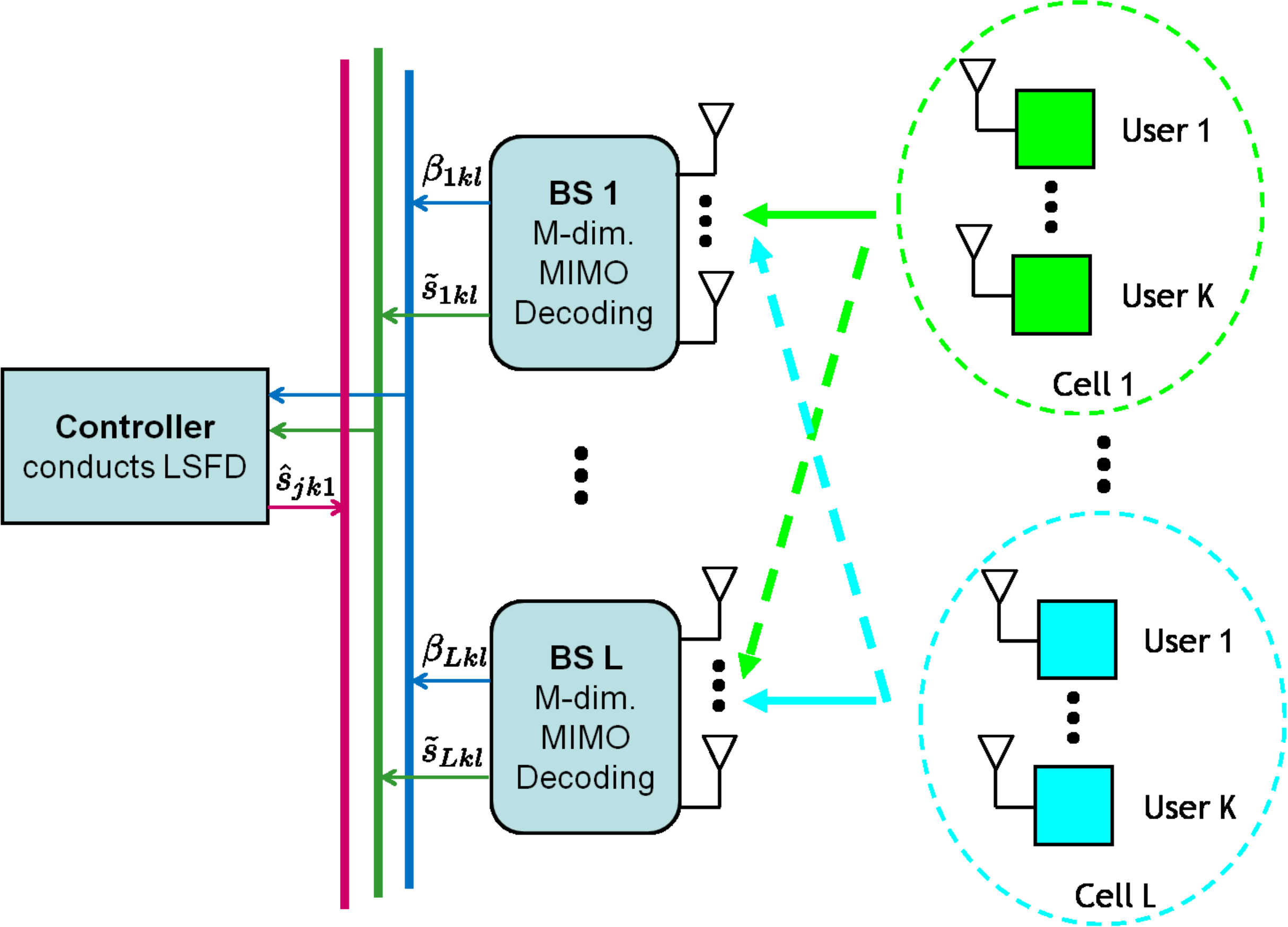}
  \caption{Block Diagram of LSFD}\label{fig:block_diag}
\end{figure}

The network architecture for this protocol is shown in Fig. \ref{fig:block_diag}.

We would like to emphasize the following points.
\begin{itemize}
\item The large scale fading coefficients $\beta_{jkl}$ are easy to estimate since they
are constant over the $M$ antennas, OFDM subcarriers, and over  $T_\beta$ OFDM symbols.
\item Steps 1 and 2 are conducted only once for every large scale coherence interval, i.e., every $T_\beta$ OFDM symbols.
\item The estimate $\hat{\gv}_{lkl}$ in Step 3 is computed once for each coherence interval, i.e., every $T$ OFDM symbols.
\item Steps 4 and 5 are conducted for each OFDM symbol.
\end{itemize}

Taking into account the above points, we conclude that in LSFD, the backhaul  traffic between base stations and the network controller grows marginally compared to the TDD protocol. Though
LSFD requires some additional computations (Steps 2 and 5) at the network controller, these computations, especially in the case of the decentralized LSFD (see Section \ref{sec:pcp-local}) when the size of matrices $\Am_k$ is small, are not overwhelming.

\subsection{LSFD with Matched Filtering $M$-dimensional Receiver}
First we assume that in Step  \ref{step:M dim decoding} of LSFD, matched filtering
(\ref{eq:stilde_kl}) is used. Let $a_{klj}$ be the $j^{\rm th}$
element of $\av_{kl}$. It is useful to represent estimates
$\hat{s}_{kl}$ as follows
\begin{align*}
\hat{s}_{kl} = \av_{kl}^\herm \tilde{\sv}_k = \sum_{j=1}^L a_{klj}^* \tilde{s}_{kj}=&\underbrace{\sum_{j=1}^L a_{klj}^* \EE [\hat{\gv}_{jkj}^\herm \gv_{jkl}] \sqrt{q_{kl}} s_{kl}}_{\rm Useful \ \ Signal}  + \underbrace{\sum_{j=1}^L a_{klj}^* \sum_{n=1,n\neq l}^L \EE [\hat{\gv}_{jkj}^\herm \gv_{jkn}] \sqrt{q_{kn}} s_{kn}}_{\rm Pilot \ \ Contamination
\ \ Interference } \nonumber\\
&+\sum_{n=1}^L \sum_{j=1}^L a_{klj}^* \left( \hat{\gv}_{jkj}^\herm
\gv_{jkn} - \EE[\hat{\gv}_{jkj}^\herm \gv_{jkn}] \right)
\sqrt{q_{kn}} s_{kn} \\
&\underbrace{+\sum_{n=1}^L \sum_{m=1,m \neq k}^K \sum_{j=1}^L a_{klj}^*
\hat{\gv}_{jkj}^\herm \gv_{jmn} \sqrt{q_{mn}} s_{mn}
+ \sum_{j=1}^L a_{klj}^* \hat{\gv}_{jkj}^\herm \zv_{j}}_{\rm Interference \ \ plus
\ \ Noise \ \ Terms}
\end{align*}

Taking into account that $s_{kl}$ and $s_{mn}$ are independent if $(k,l)\neq (m,n)$,
that $\hv_{jkl}$ are independent from $\hv_{nms}$ if $(j,k,l)\neq (n,m,s)$,
and that $\hat{\gv}_{jkl}$ are uncorrelated with $\ev_{jkl}$, it is easy to show
that all terms in the above expression are uncorrelated. Thus, we can apply
Theorem 1
from \cite{hassibi2003much}. According to this theorem, the channel that minimizes
$I(\hat{s}_{kl};s_{kl}|\hat{\gv}_{lkl})$ is the AWGN channel with noise variance
equal to the sum of the variances of interferences and noise in the above expression.
In other words we have
$
I(\hat{s}_{kl};s_{kl}|\hat{\gv}_{lkl})\ge \log(1+{\rm SINR}_{kl}),
$
where
\begin{align*}
{\rm SINR}_{kl}={\EE[|\mbox{Useful Signal}|^2]
\over \mathrm{Var}[\mbox{Pilot Cont. Interf.}]+\mathrm{Var}[\mbox{Interf. plus Noise Terms}]}
\end{align*}
This leads to the following Theorem, which was proved in \cite{ashikhminPart1}. Slightly different notations are used in
\cite{ashikhminPart1}, hence,  for the sake of self-completeness,
 we present the proof of this theorem in the Appendix.
To shorten expressions, we use the notation
\begin{equation}\label{eq:hat_a}
\hat{a}_{klj} =  \frac{\beta_{jkj} \sqrt{p_{kj}}}{1 + \sum_{i=1}^L \beta_{jki} p_{ki}} \cdot a_{klj}
\end{equation}

 \begin{thm}\label{thm:SINR for MF}
 If matched filtering decoding is used in Step \ref{step:M dim decoding} of LSFD, then the achievable ${\rm SINR}_{kl}$ for the $k^{\rm th}$ user in $l^{\rm th}$ cell is given by
 \small{
\begin{equation}\label{eq:D}
{\rm SINR}_{kl}
={\left|\sum_{j=1}^L \hat{a}_{klj}^* \beta_{jkl} \right|^2 p_{kl} q_{kl} M \over
\sum_{n=1 \atop n\neq l}^L \left|\sum_{j=1}^L \hat{a}_{klj}^* \beta_{jkn} \right|^2 p_{kn} q_{kn} M
+ \sum_{j=1}^L |\hat{a}_{klj}|^2 (1 + \sum_{i=1}^L \beta_{jki} p_{ki})
\left( 1 + \sum_{n=1}^L \sum_{m = 1}^K \beta_{jmn} q_{mn} \right)}
\end{equation}}
\end{thm}

\proof
See Appendix \ref{appendix-thm-2}.
\IEEEQED

In \cite{ashikhmin2012pilot}, the following way of LSFD, called Zero-Forcing LSFD (ZF LSFD), was proposed:
\begin{equation}\label{eq:ZF-LSFD}
\Am_k=\Bm_k^{-1}, \mbox{ and }
\Bm_k=\left(\begin{array}{ccc}
\beta_{1k1} & \ldots & \beta_{1kL}\\
\vdots & & \vdots\\
\beta_{Lk1} & \ldots & \beta_{LkL}
\end{array}\right).
\end{equation}
 It is not difficult to see that with this choice of $\Am_k$, the numerator becomes equal to $p_{ql}q_{kl}M$ and the
first term in the denominator of (\ref{eq:D}) becomes equal to zero, while its two other terms do not depend on $M$.
 Thus, we obtain  that
$$
\lim_{M\rightarrow \infty} {\rm SINR}_{kl}\stackrel{\textrm{a.s.}}{=}\infty,
$$
which is a drastic improvement over Theorem \ref{thm:finite_SINR}.

It happens,
however, that in the case of $M<10^5$, all terms in the denominator of (\ref{eq:D})
have comparable magnitudes with each other and
for getting good performance, it is not enough to cancel only the first term which is caused by the
pilot contamination interference. For this reason, ZF LSFD has very bad
performance unless $M$ is very large (see Fig.\ref{fig1}).
 The natural question is
whether LSFD can be designed so as
to mitigate the most significant interference terms of
(\ref{eq:stilde_kl}) for a given $M$. We answer positively to this question below.
 To keep notations short, we will use
$\hat{\av}_{kl}=(\hat{a}_{kl1},\dots,\hat{a}_{klL})^T$.
\begin{thm} \label{thm:optimalSINR for MF}
If matched filtering decoding is used in Step  \ref{step:M dim decoding}  of LSFD, then
the optimal LSFD coefficients $\hat{\av}_{kl}$ and the corresponding SINRs are
\begin{equation}\label{eq:opta_MF}
 \hat{\av}_{kl,opt} = \left( \sum_{n=1,n \neq l}^L \betav_{kn}
\betav_{kn}^\herm p_{kn} q_{kn} M + \Lambdam_k \right)^{-1}
\betav_{kl},
\end{equation}
\begin{align}
\label{eqn:SINR}
 {\rm SINR}_{kl,opt} =\betav_{kl}^\herm \left( \sum_{n=1,n \neq l}^L \betav_{kn}
\betav_{kn}^\herm p_{kn} q_{kn} M + \Lambdam_k \right)^{-1}
\betav_{kl} p_{kl} q_{kl} M,
\end{align}
where $\betav_{kn} = [\beta_{1kn} \ldots \beta_{Lkn}]^T$ and
$\Lambdam_{k} = {\rm diag}(\lambda_{1k},\ldots,\lambda_{Lk})$ with
\begin{equation}\label{eq:lambda_jk}
\lambda_{jk} = (1 + \sum_{i=1}^L \beta_{jki} p_{ki}) \left( 1 +
\sum_{n=1}^L \sum_{m = 1}^K \beta_{jmn} q_{mn} \right).
\end{equation}
\end{thm}

\proof  Let  $\Cm$ be a Hermitian matrix. According to the Rayleigh-Ritz theorem, see for example \cite{Horn}, the maximum of
\begin{equation}\label{eq:Ritz_theorem}
{\xv^H \uv\uv^H\xv\over \xv^H \Cm \xv}
\end{equation}
is achieved at
\begin{equation}\label{eq:xv=C^-1*uv}
\xv=\Cm^{-1}\uv
\end{equation}

After some efforts, we transform (\ref{eq:D}) into the following form
$$
{\rm SINR}_{kl}={\hat{\av}_{kl}^\herm \betav_{kl}
\betav_{kl}^\herm p_{kl} q_{kl} M \hat{\av}_{kl} \over \hat{\av}_{kl}^\herm
\left( \sum_{n=1,n \neq l}^L \betav_{kn}
\betav_{kn}^\herm p_{kn} q_{kn} M + \Lambdam_k \right) \hat{\av}_{kl}}
$$
It is easy to check that the matrix in the denominator is Hermitian. Hence we can
apply (\ref{eq:xv=C^-1*uv}). After simple computations, we obtain the assertions. \IEEEQED

It is important  to note that the vector
$\hat{\av}_{kl,opt}$ that maximizes the SINR of user $k$ in cell $l$ can
be computed independently of the other vectors $\hat{\av}_{mn,opt}$.

\subsection{LSFD with Zero-Forcing $M$-dimensional decoding}\label{subsec:LSFD_ZF}

Now let us consider the scenario where zero forcing is used as an $M$-dimensional receiver in Step  \ref{step:M dim decoding} of  LSFD.
The BS in cell $l$ conducts linear zero forcing by taking the Moore-Penrose pseudoinverse of the estimated channel matrix as
$$
\Vm_l=[\vv_{l1},\ldots,\vv_{lK}]=\hat{\Gm}_l^H(\hat{\Gm}_l\hat{\Gm}_l^H)^{-1}.
$$
and computing
\begin{equation}\label{eq:ZFDecoding}
 \tilde{s}_{kl}^{\rm}=\vv_{lk}^\herm \yv_l,
\end{equation}
where $\vv_{lk}$ denotes the $k^{\rm th}$ column of $\Vm_l$.
Therefore,
\begin{align} \label{eqn:zf-condn}
\vv_{lk}^\herm \hat{\gv}_{lml} &= 0, \ \ \forall \ \ m \neq k, \mbox{ and }  \vv_{lk}^\herm \hat{\gv}_{lkl} = 1.
\end{align}

After zero forcing for the $k^{\rm th}$ user of $l^{\rm th}$ cell, we get
\begin{align}
 \tilde{s}_{kl}^{\rm}= \vv_{lk}^\herm \yv_l =&\sum_{n=1}^L \vv_{lk}^\herm \hat{\gv}_{lkn} \sqrt{q_{kn}} s_{kn} + \sum_{n=1}^L \vv_{lk}^\herm \ev_{lkn} \sqrt{q_{kn}} s_{kn} + \sum_{n=1}^L \sum_{m \neq k} \vv_{lk}^\herm \hat{\gv}_{lmn} \sqrt{q_{mn}} s_{mn} \nonumber\\
 & + \sum_{n=1}^L \sum_{m \neq k} \vv_{lk}^\herm \ev_{lmn} \sqrt{q_{mn}} s_{mn}
 + \vv_{lk}^\herm \zv_l\nonumber\\
=\ & \vv_{lk}^\herm \hat{\gv}_{lkl} \sqrt{q_{kl}} s_{kl} + \sum_{n=1, n\neq l}^L \frac{\beta_{lkn}\sqrt{p_{kn}}}{\beta_{lkl}\sqrt{p_{kl}}}\vv_{lk}^\herm \hat{\gv}_{lkl} \sqrt{q_{kn}}
s_{kn}+\sum_{n=1}^L \sum_{m = 1}^K \vv_{lk}^\herm \ev_{lmn} \sqrt{q_{mn}} s_{mn} + \vv_{lk}^\herm \zv_l\nonumber\\
\stackrel{(a)}{=}& \underbrace{\sqrt{q_{kl}} s_{kl}}_{\rm Useful\ \ Term} + \underbrace{\sum_{n=1, n\neq l}^L \frac{\beta_{lkn}\sqrt{p_{kn}}}{\beta_{lkl}\sqrt{p_{kl}}} \sqrt{q_{kn}} s_{kn}}_{\rm Pilot Contamination Term} + \underbrace{\sum_{n=1}^L \sum_{m = 1}^K \vv_{lk}^\herm \ev_{lmn} \sqrt{q_{mn}} s_{mn} + \vv_{lk}^\herm \zv_l}_{\rm Interference \ \ + \ \ Noise\ \ Terms},
 \end{align}
 where $(a)$ follows from (\ref{eqn:zf-condn}).

After LSFD, following the same steps as in Section \ref{sec:analysis-pcp}, we get
\begin{align}\label{eq:hat_skl_zero_forcing_receiver}
\hat{s}_{kl} =\av_{kl}^\herm \tilde{\sv}_k = \sum_{j=1}^L a_{klj}^* \tilde{s}_{kj} =& \underbrace{\sum_{j=1}^L a_{klj}^* \frac{\beta_{jkl}\sqrt{p_{kl}}}{\beta_{jkj}\sqrt{p_{kj}}} \sqrt{q_{kl}} s_{kl}}_{\rm Useful \ \ Signal} + \underbrace{\sum_{n=1,n \neq l}^L \sum_{j=1}^L a_{klj}^* \frac{\beta_{jkn}\sqrt{p_{kn}}}{\beta_{jkj}\sqrt{p_{kj}}} \sqrt{q_{kn}} s_{kn}}_{\rm Pilot \ \ Contamination}\nonumber\\
&+ \underbrace{\sum_{n=1}^L \sum_{m = 1}^K \sum_{j=1}^L a_{klj}^* \vv_{jk}^\herm \ev_{jmn} \sqrt{q_{mn}} s_{mn} + \sum_{j=1}^L a_{klj}^* \vv_{jk}^\herm \zv_{j}}_{\rm Interference~+~ Noise \ \ Terms}
\end{align}

The variances of terms in the denominator can be found as follows. Since $s_{kn}$ and $s_{jl}$ are independent if $(k,n)\not =(j,l)$ we have
\begin{align*}
\mathrm{Var}\left[ |{\rm Pilot \ \ Contamination}|^2   \right] = &\sum_{n=1,n\neq l}^L \left|\sum_{j=1}^L a_{klj}^* \frac{\beta_{jkn}\sqrt{p_{kn}}}{\beta_{jkj}\sqrt{p_{kj}}} \sqrt{q_{kn}} \right|^2 \EE [| s_{kn}|^2] \nonumber\\
= &\sum_{n=1,n\neq l}^L \left|\sum_{j=1}^L a_{klj}^* \frac{\beta_{jkn}\sqrt{p_{kn}}}{\beta_{jkj}\sqrt{p_{kj}}} \right|^2 q_{kn}
\end{align*}
Since $s_{kn}$ are independent from all $\zv_{j}$, and $\zv_{i}$ is independent from $\zv_{j}$ if $i\not =j$, we have
\begin{align*}
&\mathrm{Var}\left[ |{\rm Interference+Noise Terms}|^2 | \beta_{lkn}, l,n=1,L \} \right] \nonumber\\
= &\sum_{n=1}^L \sum_{m = 1}^K \EE \left[ \left| \sum_{j=1}^L a_{klj}^* \vv_{jk}^\herm \ev_{jmn} \right|^2 \right] q_{mn} \EE [ |s_{mn}|^2 ] +\EE \left[ \left| \sum_{j=1}^L a_{klj}^* \vv_{jk}^\herm \zv_{j} \right|^2 \right] \nonumber\\
=& \sum_{n=1}^L \sum_{m = 1}^K \sum_{j=1}^L |a_{klj}|^2 \EE \left[ \vv_{jk}^\herm \ev_{jmn} \ev_{jmn}^\herm \vv_{jk} \right] q_{mn} + \sum_{j=1}^L |a_{klj}|^2 \EE \left[ \vv_{jkj}^\herm \zv_{j} \zv_{j}^\herm \vv_{jkj} \right] \nonumber\\
=& \sum_{n=1}^L \sum_{m = 1}^K \sum_{j=1}^L |a_{klj}|^2 \EE_{\vv_{jk}}[\vv_{jk}^\herm \vv_{jk}] \times \left( \beta_{jmn} - \frac{\beta_{jmn}^2 p_{mn}}{1 + \sum_{i=1}^L \beta_{jmi}p_{mi}}\right) q_{mn} + \sum_{j=1}^L |a_{klj}|^2 \EE_{\vv_{jk}}[\vv_{jk}^\herm \vv_{jk}] \nonumber\\
=& \sum_{j=1}^L |a_{klj}|^2 \EE_{\vv_{jk}}[\vv_{jk}^\herm \vv_{jk}] \times \left[ \sum_{n=1}^L \sum_{m = 1}^K  \left( \beta_{jmn} - \frac{\beta_{jmn}^2 p_{mn}}{1 + \sum_{i=1}^L \beta_{jmi}p_{mi}}\right) q_{mn} + 1 \right]
\end{align*}
Using standard result from random matrix theory \cite{tulino2004random}, we obtain
$$\EE_{\vv_{jk}} [\vv_{jk}^\herm \vv_{jk}] = \frac{1 + \sum_{i=1}^L \beta_{jki} p_{ki}}{\beta_{jkj}^2 p_{kj} (M - K)}.$$

All terms in (\ref{eq:hat_skl_zero_forcing_receiver}) are
uncorrelated.   Thus, according to \cite{hassibi2003much}, the worst case channel is the AWGN channel
 with variance equal to the sum of the interferences and noise variances found above.
Using
$\hat{a}_{klj} = \frac{a_{klj}}{\beta_{jkj} \sqrt{p_{kj}}}
$, after some computations,  we get the following result.
\begin{thm}\label{thm:SINR for ZF} If zero-forcing decoding is used in Step  \ref{step:M dim decoding}  of LSFD, then the achievable SINRs
are
\small{
$$
{\rm SINR}_{kl} = \frac{\left|\sum_{j=1}^L \hat{a}_{klj}^* \beta_{jkl} \right|^2 p_{kl} q_{kl}}{\begin{split} &\sum_{n=1,n\neq l}^L \left|\sum_{j=1}^L \hat{a}_{klj}^* \beta_{jkn} \right|^2 p_{kn} q_{kn}
 \\&+ \sum_{j=1}^L \frac{|\hat{a}_{klj}|^2}{M-K} (1 + \sum_{i=1}^L \beta_{jki} p_{ki})
  \left( 1 + \sum_{n=1}^L \sum_{m = 1}^K \left[ \beta_{jmn} - \frac{\beta_{jmn}^2 p_{mn}}{1 + \sum_{i=1}^L \beta_{jmi}p_{mi}} \right] q_{mn} \right)\end{split}}
$$}

\end{thm}
 Let us define diagonal matrices
 $\Lambdam_{k} = {\rm diag}(\lambda_{1k},\lambda_{2k},\ldots,\lambda_{Lk})$
 with
\begin{align}\label{eq:lambda_jk_zf}
{\lambda}_{jk} = \frac{M}{M-K} \left( 1 + \sum_{i=1}^L \beta_{jki} p_{ki} \right) \times \left( 1 + \sum_{n=1}^L \sum_{m = 1}^K \left[ \beta_{jmn} -
\frac{\beta_{jmn}^2 p_{mn}}{1 + \sum_{i=1}^L \beta_{jmi}p_{mi}}
\right] q_{mn} \right)
\end{align}
 and let again $\betav_{kn} = (\beta_{1kn} \beta_{1kn} \ldots \beta_{Lkn})^T$.
 With these notations, we have the following theorem.

\begin{thm}\label{thm:optimalSINR for ZF} If zero-forcing decoding is used in Step  \ref{step:M dim decoding}  of LSFD,
then the
optimal  $\hat{a}_{kl,opt}$ and the corresponding achievable SINRs are defined
by (\ref{eq:opta_MF}) and (\ref{eqn:SINR}) with ${\lambda}_{jk}$ defined in (\ref{eq:lambda_jk_zf}).
\end{thm}
A proof of this result is similar to the proof of Theorem \ref{thm:optimalSINR for MF}.

\section{Transmit Power Optimization} \label{sec:pcp-optz}

Using optimal LSFD coefficients obtained in Theorems \ref{thm:optimalSINR for MF}  and \ref{thm:SINR for ZF} already give significant improvement compared to the case when LSFD is not used.
However, even greater improvements can be obtained if we optimize the transmit powers $p_{kl}$ and $q_{kl}$.
Denote
$$
\pvx=(p_{kl}:~k=1,K,~l=1,L), \mbox{ and } \qv=(q_{kl}:~k=1,K,~l=1,L).
$$
In this work, we will assume constant powers $\pvx=P_{\max}\onev$  during the training phase
and focus on optimization of transmit powers $\qv$ during the data transmission phase.
As indicated in Section \ref{sec:intro}, we shall optimize system performance with respect to the max-min criterion
\begin{align} \label{eq:optz-q}
\max_{\qv} \min_{k,l} & \ \ ~P_{\max} M q_{kl} \cdot \betav_{kl}^\herm \left( \sum_{n=1,n \neq l}^L \betav_{kn} \betav_{kn}^\herm q_{kn}P_{\max} M + \Lambdam_k \right)^{-1} \betav_{kl}  \nonumber\\
\mbox{ subject  to } &\ \  ~\zerov \leq \qv \leq Q_{\max} \onev,
\end{align}
where $\onev$ is the $KL \times 1$ all ones vector, and $\Lambdam_k$ is defined in (\ref{eq:lambda_jk}) with $p_{ki}=P_{\max},\forall k,i$.
Let
\begin{align*}
\gamma  \doteq \min_{k,l}~ \ \betav_{kl}^\herm \left( \sum_{n=1,n \neq l}^L \betav_{kn} \betav_{kn}^\herm P_{\max} q_{kn} M + \Lambdam_k \right)^{-1} \betav_{kl} \times P_{\max} q_{kl} M
\end{align*}
It is convenient to reformulate the optimization problem  (\ref{eq:optz-q}) in the following form
\begin{align}
\max_{\qv} &\ \ ~\gamma \nonumber\\
{\rm subject\ to} &\ \  \zerov \leq \qv \leq Q_{\max} \onev, \nonumber\\
&\ \ P_{\max}M q_{kl} \cdot \betav_{kl}^\herm \left( \sum_{n=1,n \neq l}^L \betav_{kn} \betav_{kn}^\herm p_{kn} q_{kn} M + \Lambdam_k \right)^{-1} \betav_{kl}   \geq \gamma,\forall k,l. \label{eqn:optz-q-2}
\end{align}
We solve the problem (\ref{eqn:optz-q-2}) in an iterative fashion. We start with an initial value of $\gamma = \frac{\gamma_{\max} + \gamma_{\min}}{2}$ where $\gamma_{\max}$ and $\gamma_{\min}$ are chosen apriori and follow the bisection algorithm until the difference between $\gamma_{\max}$ and $\gamma_{\min}$ becomes small. The algorithm can be summarized as follows.

\noindent{{\bf Optimal Power Allocation} \\
\noindent {\bf Input:}
$P_{\max},Q_{\max},\beta_{jkl},~j,l=1,L;k=1,K$.\\
\noindent {\bf Output:}
$\gamma^{opt}=\max_{0\le \qv\le Q_{\max}\onev} \min_{k,l} {\rm SINR}_{kl};~q_{kl}^{opt}, \forall k,l$.
\begin{enumerate}
\item {\bf Step 1}: Set $\gamma_{\max} = \max_{k,l} || \betav_{kl} ||^2 P_{\max} Q_{\max} M$ and $\gamma_{\min} = 0$.
\item {\bf Step 2}: Assign  $\gamma = \frac{\gamma_{\max} + \gamma_{\min}}{2}$.
\item {\bf Step 3}: Check feasibility of the following problem
\begin{align}
\min \ \ &\sum_{k=1}^K \sum_{l=1}^L q_{kl} \label{eqn:optz-q-3}\\
{\rm subject\ to} \ \ & \zerov \leq \qv \leq Q_{\max} \onev, \nonumber\\
& p_{kl} q_{kl} M\cdot \betav_{kl}^\herm \left( \sum_{n=1,n \neq l}^L \betav_{kn} \betav_{kn}^\herm p_{kn} q_{kn} M + \Lambdam_k \right)^{-1} \betav_{kl} \geq \gamma,~k=1,K; l=1,L. \nonumber
\end{align}
\item {\bf Step 4}: If $\gamma$ is feasible, set $\gamma_{\min} = \gamma$ and go to Step 5, else set $\gamma_{\max} = \gamma$.
\item {\bf Step 5}: If $\gamma_{\max} - \gamma_{\min} < \epsilon$, where $\epsilon$ is a small number, stop and assign $\gamma^{opt}=\gamma_{\min}$ and $q_{kl}^{opt}=q_{kl}$, where $q_{kl}$ are solutions of (\ref{eqn:optz-q-3}) with $\gamma=\gamma_{\min}$, otherwise go to Step 2.
\end{enumerate}
\noindent{{\bf The end}}

Using the techniques of \cite{yates1995framework}, \cite{ulukus1997adaptive}, we show in Section \ref{sec:pcp-local}
that if the problem (\ref{eqn:optz-q-3}) is feasible, then it has a unique solution and that there are iterative algorithms that converge to it.

The optimality of the proposed algorithm can be proved by contradiction. Let the solution obtained by the algorithm is $\gamma_1$, and  the optimal solution is $\gamma_2 > \gamma_1$. Then there exists $\gamma_2 > \gamma_3 > \gamma_1$ such that $\gamma_3$ is infeasible (else our algorithm would have returned $\gamma_3$ as the optimal solution.). Since $\gamma_2 > \gamma_3$, we can reduce one user's power to make its SINR equal to $\gamma_3$. This results in a reduction of interference to all the other users, making the SINR of the other users $\geq \gamma_2 > \gamma_3$. This means $\gamma_3$ is also feasible, which is a contradiction.

The key step of the algorithm is Step 3. In Section \ref{sec:pcp-local} we propose a nice decentralized algorithm for implementing it. Now we can formulate the following communication protocol.


\noindent{\bf An LSFD with Transmit Power Optimization}
\begin{enumerate}
\item All $L$ base stations estimate their large scale fading coefficients (the $j$-th base station estimates the coefficients $\beta_{jkl},k=1,K;l=1,L$)
and send them to a controller.
\item The controller runs the {\bf Optimal Power Allocation} algorithm.
\item The controller sends the optimal transmit powers $q_{kl}^{opt}$ to the corresponding users (perhaps via the corresponding Base Stations).
\item The users transmit data with $q_{kl}^{opt}$.
\item The controller runs an LSFD  to get estimates $\hat{s}_{kl}$.
\end{enumerate}
\noindent{\bf The end.}

Simulation results in Section \ref{sec:results} show that power optimization gives large performance gain.

\begin{rem}
Note that the problem (\ref{eq:optz-q}) can also be formulated as a power optimization problem over the user powers during the training phase while keeping the powers in the data transmission phase fixed.
This problem can be solved using the same techniques described in Section \ref{sec:pcp-optz}.
\end{rem}

\section{Decentralized LSFD} \label{sec:pcp-local}
The assumption that the network controller coordinates all base stations across
the entire network is reasonable only for small networks, like a network for a campus, stadium, small town or similar facility.
In a large network, we have to use decentralized algorithms and protocols that require coordination of only a small number of BSs.
In this section, we propose  a decentralized  version of LSFD. We assume
that the $l^{\rm th}$ BS has access  only to its $L'$ neighboring
cells. Let
 $$
 \Omega^{(l)}=\left\{l\cup \{\mbox{indices of $L'$ neighboring cells of cell $l$}\}\right\}, \mbox{ and }
 \hat{L}=L'+1=|\Omega^{(l)}|.
 $$
 To make the description of the following protocol short, it
  will be convenient to assume that the $l^{\rm th}$ BS plays the role of the network controller
 for the network formed by the cells from $\Omega^{(l)}$ (a number of other possibilities for organizing a network controller or controllers exist). We  assume for each $l$, the elements of $\Omega^{(l)}$ are ordered in a certain order and the order is fixed.

\noindent{\bf Decentralized Uplink LSFD}
\begin{enumerate}
\item\label{step:eta}
 The $l^{\rm th}$ BS estimates $\beta_{lkn},n=1,L$, and computes
\begin{equation}\label{eq:eta}
\eta_{kl}={\beta_{lkl}\sqrt{p_{kl}}\over 1+\sum_{i=1}^L
\beta_{lki}p_{ki} },k=1,K.
\end{equation}
(See notes at the end of this section on an empirical way of  computation of $\eta_{kl}$.)
\item\label{step:Mdim_decoding_decPCP} The $l^{\rm th}$ BS computes $\tilde{s}_{kl},k=1,K$, using
an $M$-dimensional decoding procedure. In particular, it may apply matched filtering (\ref{eq:stilde_kl}) or zero-forcing
(\ref{eq:ZFDecoding}) decoding.
\item \label{step:stilde} The $l^{\rm th}$ BS collects from neighboring BSs symbols $\tilde{s}_{kj},j\in \Omega^{(l)}$,
 and forms the vectors
\begin{equation}\label{eq:s_k^(l)}
\tilde{\sv}^{(l)}_k = [\tilde{s}_{kj}: j\in \Omega^{(l)}]^T,k=1,K.
\end{equation}
 \item\label{step:a_kl,dec} The  $l^{\rm th}$ BS collects coefficients $\eta_{kj},j\in \Omega^{(l)}$, and computes
 $\hat{L}$-dimensional vectors $\av_{kl,{dec}}=(a_{klj,dec}:j\in \Omega^{(l)})$ (see explanations
below). Here `${dec}$' stands for `decentralized'.
\item\label{step:hats_kl}
 The $l^{\rm th}$ BS
computes the LSFD estimates as $ \hat{s}_{kl,{dec}}=\av_{kl,{dec}}^\herm \tilde{\sv}^{(l)}_k,k=1,K$.
\end{enumerate}
\noindent{\bf The end.}

\subsection{Decentralized LSFD with Matched Filtering $M$-dimensional Receiver}\label{subsec:DecLSFD_MF}
If  we use matched filtering in Step  \ref{step:Mdim_decoding_decPCP} of LSFD, we get
\begin{align}
\hat{s}_{kl,{dec}}=\ & \av_{kl,{dec}}^\herm \tilde{\sv}^{(l)}_k = \sum_{j\in \Omega^{(l)}}
a_{klj,{dec}}^* \tilde{s}_{kj} \nonumber\\
=& \underbrace{\sum_{j\in \Omega^{(l)}}
 a_{klj,{dec}}^* \EE[\hat{\gv}_{jkj}^\herm \gv_{jkl}] \sqrt{q_{kl}} s_{kl}}_{\rm Useful \ \ Signal} +  \underbrace{\sum_{j\in \Omega^{(l)}} a_{klj,{dec}}^* \sum_{n=1,n\neq l}^L \EE[\hat{\gv}_{jkj}^\herm \gv_{jkn}] \sqrt{q_{kn}} s_{kn}}_{\rm Pilot \ \ Contamination}\nonumber\\
&+ \sum_{n=1}^L \sum_{j\in \Omega^{(l)}} a_{klj,{dec}}^* \left(
\hat{\gv}_{jkj}^\herm \gv_{jkn} - \EE[\hat{\gv}_{jkj}^\herm
\gv_{jkn}] \right) \sqrt{q_{kn}} s_{kn} \nonumber\\
&+ \underbrace{\sum_{n=1}^L \sum_{m \neq k} \sum_{j\in \Omega^{(l)}} a_{klj,{dec}}^* \hat{\gv}_{jkj}^\herm \gv_{jmn} \sqrt{q_{mn}} s_{mn} + \sum_{j\in \Omega^{(l)}} a_{klj,{dec}}^*
\hat{\gv}_{jkj}^\herm \zv_{j}}_{\rm
Interference \ \ plus \ \ Noise \ \ Terms} \label{eq:s_kl,dec}
\end{align}

Let us, similar to (\ref{eq:hat_a}), define
$\hat{a}_{klj,dec}=\eta_{kj}\cdot a_{klj,dec}$.
Conducting derivations similar to the ones used in Theorem \ref{thm:SINR for MF}, we obtain the following
expression
{\small
\begin{align}\label{eq:decPCP_MF}
&{\rm SINR}_{kl,dec}\nonumber\\
 = &\frac{Mp_{kl} q_{kl} \cdot \left| \sum_{j\in \Omega^{(l)}} \hat{a}_{klj,dec}^* \beta_{jkl} \right|^2}
{M\cdot\sum_{n=1,n\neq l}^L p_{kn} q_{kn}\left|  \sum_{j\in \Omega^{(l)}} \hat{a}_{klj,dec}^* \beta_{jkn} \right|^2
 + \sum_{j\in\Omega^{(l)}} |\hat{a}_{klj,dec}|^2 (1 + \sum_{i=1}^L \beta_{jki} p_{ki}) \left( 1 + \sum_{n=1}^L \sum_{m = 1}^K \beta_{jmn} q_{mn} \right)}
\end{align}
}

Further, by defining vectors $\betav_{kn}^{(l)} = (\beta_{jkn}: j\in \Omega^{(l)})^T$ and using arguments similar to the  ones used in
Theorem \ref{thm:optimalSINR for MF}, we conclude that ${\rm SINR}_{kl,dec}$ is maximized at
\begin{equation}\label{eq:a_kl optimal Decetr LSFD}
\hat{\av}_{kl,opt,dec} = \left( \sum_{n=1,n \neq l}^L \betav^{(l)}_{kn} \betav_{kn}^{{(l)} \herm} p_{kn} q_{kn} M + \Lambdam^{(l)}_k \right)^{-1} \betav^{(l)}_{kl},
\end{equation}
where
$\Lambdam^{(l)}_{k} = {\rm diag}(\lambda_{jk}:j\in \Omega^{(l)} )$ and
\begin{equation}\label{eq:lambda}
\lambda_{jk} = (1 + \sum_{i=1}^L \beta_{jki} p_{ki}) \left( 1 + \sum_{n=1}^L \sum_{m = 1}^K \beta_{jmn} q_{mn} \right)
\end{equation}

Let
$\Dm_k ={\rm diag} \left( \eta_{kj}: j\in \Omega^{(l)}\right)$.
Coefficients $\eta_{kj},j\in \Omega^{(l)}$,  are passed to the $l^{\rm th}$ BS in Step \ref{step:a_kl,dec}. So, if the $l^{\rm th}$ BS possesses
$\hat{\av}_{kl,opt,dec}$, it
could compute
${\av}_{kl,opt,dec} =\Dm_k^{-1}\hat{\av}_{kl,opt,dec}$ and use it in Step \ref{step:hats_kl} of the algorithm.
The problem is, however, that the $l^{\rm th}$ BS does not know the powers $p_{kn}$ and $q_{kn}$ for $n\not \in \Omega^{(l)}$.
Thus it can not compute optimal $\hat{\av}_{kl,dec,opt}$ according to
(\ref{eq:a_kl optimal Decetr LSFD}).
Using in (\ref{eq:a_kl optimal Decetr LSFD}), for instance, maximum powers $p_{kn}=P_{max}$ and $q_{kn}=Q_{max}$ or minimum
powers $p_{kn}=0$ and $q_{kn}=0$ for $n\not\in
\Omega^{(l)}$ results in significant performance degradation.

To resolve this problem, nstead of computing $\hat{\av}_{kl,opt,dec}$ according to
(\ref{eq:a_kl optimal Decetr LSFD}), we propose that the $l^{\rm th}$ BS empirically  estimates the matrix $\EE \left[ \tilde{\sv}^{(l)}_k \tilde{\sv}^{(l)^\herm}_k \right],$ and further computes
\begin{align*}
{\av}_{kl}^{\rm MMSE}&= \EE \left[ \tilde{\sv}^{(l)}_k \tilde{\sv}^{(l)^\herm}_k \right]^{-1} \EE \left[ \tilde{\sv}^{(l)}_k s_{kl}^* \right]
\end{align*}
It is shown in the next theorem that vector $\EE \left[ \tilde{\sv}^{(l)}_k s_{kl}^* \right]$ can be computed by the $l^{\rm th}$ BS directly.
Estimation of matrix $\EE \left[ \tilde{\sv}^{(l)}_k \tilde{\sv}^{(l)^\herm}_k \right]$ can be obtained, for instance, by
  collecting  sufficiently many samples of $\tilde{\sv}^{(l)}_k$  to get an estimate of the matrix,
and by further updating the matrix  using $\tilde{\sv}^{(l)}_k$ obtained in
Step \ref{step:stilde}.
The following theorem shows that the vectors ${\av}_{kl}^{\rm MMSE}$ are optimal.

\begin{thm} If matched filtering decoding is used in Step  \ref{step:M dim decoding} of LSFD, then
vectors
\begin{align*}
{\av}_{kl}^{\rm MMSE}&=\EE \left[ \tilde{\sv}^{(l)}_k \tilde{\sv}^{(l)^\herm}_k \right]^{-1} \EE \left[ \tilde{\sv}^{(l)}_k s_{kl}^* \right] = \Dm_k^{-1} \times const\times \hat{\av}_{kl,opt,dec},
\end{align*}
are optimal and lead to the optimal value
\begin{align} \label{eqn:SINR-limited}
{\rm SINR}_{kl,opt,dec} =  p_{kl} q_{kl} M\cdot \betav_{kl}^{(l) \herm} \left( \sum_{n=1,n \neq l}^L \betav^{(l)}_{kn} \betav_{kn}^{{(l)} \herm} p_{kn} q_{kn} M + \Lambdam^{(l)}_k \right)^{-1} \betav^{(l)}_{kl}.
\end{align}
\end{thm}
\proof
To simplify notations,
let us denote
\begin{align*}
\xv&=(x_1,\ldots,x_{\hat{L}})^T=(\hat{\gv}_{jkj}^\herm \gv_{jkl}:j\in \Omega^{(l)})^T,~\hat{\xv}=(\hat{x}_1,\ldots,\hat{x}_{\hat{L}})^T=
(\hat{\gv}_{jkj}^\herm \hat{\gv}_{jkl}:j\in \Omega^{(l)})^T \mbox{ and }\\
\bvx&=\left(b_1, \ldots, b_{\hat{L}}\right)=(\frac{\beta_{jkl} \sqrt{p_{kl}}}{\beta_{jkj} \sqrt{p_{kj}}}:j\in \Omega^{(l)}).
\end{align*}
Then we have
\begin{align}
&\EE \left[ \tilde{\sv}^{(l)}_k s_{kl}^* \right] =  \EE[\xv]
 \sqrt{q_{kl}}
= \begin{pmatrix}
b_1 \EE[\hat{x}_1] \\
\vdots\\
b_{\hat{L}} \EE[\hat{x}_{\hat{L}}] \\
\end{pmatrix}\sqrt{q_{kl}}
= \left(\frac{\beta_{jkl} \sqrt{p_{kl}}\beta_{jkj} \sqrt{p_{kj}}}{1 + \sum_{i=1}^L \beta_{jki} p_{ki}}:j\in \Omega^{(l)}\right)^T
M \sqrt{q_{kl}} \nonumber \\
&= M \Dm_k \betav^{(l)}_{kl} \sqrt{q_{kl}} \sqrt{p_{kl}}.
\end{align}
All components of this equation are available to the $l^{\rm th}$ BS, so it can compute this vector directly.

It is not difficult to show that if $(n,m,t)\not =(j,k,l)$, then
\begin{align}\label{eq:E[hhhh]}
\EE\left[ |\gv_{jkl}^H\gv_{jkl}|^2\right] &=\beta_{jkl}^2(M^2+M) \mbox{ and }
\EE\left[ |\gv_{jkl}^H\gv_{nmt}|^2\right] =\beta_{jkl}\beta_{nmt}M
\end{align}
Using (\ref{eq:s_k^(l)}) and (\ref{eq:s_tilde0}), and further (\ref{eq:E[hhhh]}), after some efforts (we omit tedious computations), we get
$$
\EE \left[ \tilde{\sv}^{(l)}_k \tilde{\sv}^{(l) \herm}_k \right] = M^2 \sum_{n=1}^L \Dm_k \betav^{(l)}_{kn}
 \betav_{kn}^{{(l)} \herm} p_{kn}q_{kn} \Dm_k + M \Dm_k \Lambdam^{(l)}_k \Dm_k
$$
As a result, we have
\begin{align}\label{eq:aMMSE}
{\av}_{kl}^{\rm MMSE} =& \left[ M^2 \sum_{n=1}^L \Dm_k \betav^{(l)}_{kn} \betav_{kn}^{{(l)}\herm}  p_{kn}q_{kn} \Dm_k + M \Dm_k \Lambdam^{(l)}_k \Dm_k \right]^{-1} M \Dm_k \betav^{(l)}_{kl} \sqrt{q_{kl}} \sqrt{p_{kl}}\nonumber\\
=\ & \Dm_k^{-1} \left[ M \sum_{n=1}^L \betav^{(l)}_{kn} \betav_{kn}^{{(l)} \herm}p_{kn}q_{kn} + \Lambdam^{(l)}_k \right]^{-1} \betav^{(l)}_{kl} \sqrt{q_{kl}} \sqrt{p_{kl}} \nonumber\\
\stackrel{(a)}{=}\ & \frac{\sqrt{q_{kl}} \sqrt{p_{kl}}}{1 + \betav^{(l) \herm}_{kl} \left[ M \sum_{n=1 , n \neq l}^L \betav^{(l)}_{kn} \betav_{kn}^{^{(l)} \herm}p_{kn}q_{kn} + \Lambdam^{(l)}_k \right]^{-1} \betav^{(l)}_{kl}}  \nonumber\\
&\times \Dm_k^{-1} \left[ M \sum_{n=1 , n \neq l}^L \betav^{(l)}_{kn} \betav_{kn}^{{(l)} \herm} + \Lambdam^{(l)}_k \right]^{-1} \betav^{(l)}_{kl}
= const \times \Dm_k^{-1} \times  \hat{\av}_{kl,opt,dec},
\end{align}
where $(a)$ is due to the fact that
$$(\Km + \xv \xv^\herm)^{-1} \xv = \left[ \Km^{-1} - \frac{\Km^{-1} \xv \xv^\herm \Km^{-1}}{1 + \xv^\herm \Km^{-1} \xv} \right] \xv = \frac{1}{1 + \xv^\herm \Km^{-1} \xv} \Km^{-1} \xv.$$

From (\ref{eq:decPCP_MF}), it follows that vectors
$const\times \hat{\av}_{kl}$ lead to the same $SINR_{kl,dec}$ as vectors $\hat{\av}_{kl}$. Hence, vectors ${\av}_{kl}^{\rm MMSE}$ are optimal and being used in Decentralized Uplink LSFD allow achieving ${\rm SINR}_{kl,opt,dec} $ defined in
(\ref{eqn:SINR-limited}).
\IEEEQED

Now, let us return to the computation of coefficients $\eta_{kj}$ defined in Step \ref{step:eta} of the algorithm. If all users use the same training powers, i.e., $p_{kl}=p, \forall k,l$, and we assume that the $j^{\rm th}$ BS knows all $\beta_{jkl}, \forall k,l$, i.e.,  all the large scale fading coefficients between itself and all users across the entire network, then coefficients $\eta_{kj}$ can be computed directly. If users use different
training powers $p_{kl}$, then $\eta_{kj}$ can be computed empirically as follows. According to (\ref{eq:hhat_lkl}),
$$
\eta_{kj}={\EE[\gv_{lkl}^\herm \rv_{kl}] \over
\EE[\rv_{kl}^\herm \rv_{kl}]}.
$$
The quantity $\EE[\gv_{lkl}^\herm \rv_{kl}]= M \beta_{jkj}\sqrt{p_{kj}}$ can be computed directly, and $\EE[\rv_{kl}^\herm \rv_{kl}]$ can be computed empirically over multiple realizations of $\rv_{kl}$.

\subsection{Decentralized LSFD with Zero-Forcing $M$-dimensional decoding}
Let now  $M$-dimensional zero-forcing decoding is used.
Similar to the previous subsection, let $\Lambda_k^{(l)}=\diag(\lambda_{jk}:j\in \Omega^{(l)})$
 but with $\lambda_{jk}$ defined by (\ref{eq:lambda_jk_zf}). Using arguments similar to the ones used in Sections \ref{subsec:LSFD_ZF} and \ref{subsec:DecLSFD_MF}, we obtain that the SINR value is defined in (\ref{eq:SINR_Dec_ZF}), where
 $$\hat{a}_{klj,dec} = \frac{a_{klj,dec}}{\beta_{jkj} \sqrt{p_{kj}}}.
$$
\begin{equation}\label{eq:SINR_Dec_ZF}
\small{
{\rm SINR}_{kl,dec} = \frac{\left|\sum_{j=1}^L \hat{a}_{klj,dec}^* \beta_{jkl} \right|^2 p_{kl} q_{kl}}{\begin{split} &\sum_{n=1,n\neq l}^L \left|\sum_{j=1}^L \hat{a}_{klj,dec}^* \beta_{jkn} \right|^2 p_{kn} q_{kn} \\
 &+ \sum_{j=1}^L \frac{|\hat{a}_{klj,dec}|^2}{M-K} (1 + \sum_{i=1}^L \beta_{jki} p_{ki})
 \left( 1 + \sum_{n=1}^L \sum_{m = 1}^K \left[ \beta_{jmn} - \frac{\beta_{jmn}^2 p_{mn}}{1 + \sum_{i=1}^L \beta_{jmi}p_{mi}} \right] q_{mn} \right)\end{split}} }
\end{equation}
Combining arguments of
Sections \ref{subsec:LSFD_ZF} and \ref{subsec:DecLSFD_MF}, we obtain that optimal $\hat{a}_{kl,opt,dec}$ and corresponding ${\rm SINR}_{kl,opt,dec}$ are defined by
(\ref{eq:a_kl optimal Decetr LSFD}) and (\ref{eqn:SINR-limited}) respectively with $\lambda_{jk}$ defined in (\ref{eq:lambda_jk_zf}).
The $l^{\rm th}$ BS can not directly compute $\hat{{\av}}_{kl,opt,dec}$. Instead, it should empirically estimate the vector
\begin{align*}
{\av}_{kl}^{\rm MMSE}&= \EE \left[ \tilde{\sv}^{(l)}_k \tilde{\sv}^{(l)^\herm}_k \right]^{-1} \EE \left[ \tilde{\sv}^{(l)}_k s_{kl}^* \right],
\end{align*}
where vectors $\tilde{\sv}^{(l)}$ are obtained in Step \ref{step:stilde} with $M$-dimensional  zero-forcing  decoding.

Simulation results (see Section \ref{sec:results}) show that Decentralized LSFD with $M$-dimensional  zero-forcing  decoding visibly outperforms the one with $M$-dimensional
matched filtering decoding.

\subsection{Decenralized LSFD with Transmit Power Optimization}\label{subsec:LSFD_with_power_opt}

The performance of Decenralized LSFD can be significantly enhanced by choosing optimal transmit powers.
 We formulate the following optimization problem
\begin{eqnarray} \label{eqn:optz-q-limited}
\max_{\qv} && \gamma \nonumber\\
{\rm subject\ to} && \zerov \leq \qv \leq Q_{\max} \onev \mbox{ and }
{\rm SINR}_{kl,{dec}} \geq \gamma,~k=1,K,~l=1,L.
\end{eqnarray}
This optimization problem cannot be solved in
a centralized manner, but we can solve the following
optimization problem in a distributed manner for a given target SINR
$\gamma$ \cite{yates1995framework}, i.e.,
\begin{eqnarray} \label{eqn:optz-q-3-limited}
\min && \sum_{k=1}^K \sum_{l=1}^L q_{kl} \nonumber\\
{\rm subject\ to} && \zerov \leq \qv \leq Q_{\max} \onev \mbox{ and }
{\rm SINR}_{kl,{dec}} \geq \gamma,~k=1,K,~l=1,L.
\end{eqnarray}

\noindent{{\bf Decentralized LSFD with Transmit Power Optimization} }
\begin{enumerate}
\item Identical to Step 1 of {\bf Decentralized LSFD}.
\item Assign $n=1$ and repeat steps \ref{step:loop_begins}-\ref{step:loop_ends} until $|{\rm SINR}_{kj}-\gamma|<\epsilon,k=1,K;j=1,L$.
\item\label{step:loop_begins} The $l^{\rm th}$ BS computes $\tilde{s}_{kl},k=1,K$, with
matched filtering (\ref{eq:stilde_kl}) or zero-forcing
(\ref{eq:ZFDecoding}).
\item The $l^{\rm th}$ BS collects signals $\tilde{s}_{kj}$ for $j \in \Omega^{(l)}\setminus l$,
 $k=1,K$, forms
the vectors
$\tilde{\bf s}_{k}=[\tilde{s}_{kl},l\in \Omega^{(l)}_j]$,
and estimates $\EE \left[ \tilde{\sv}^{(l)}_k \tilde{\sv}^{(l)^\herm}_k \right]^{-1}$ (over multiple realizations of $\tilde{\bf s}_{k}$).
\item The $l^{\rm th}$ BS computes
$
\hat{\av}_{kl,dec}=\hat{\av}_{kl}^{\rm MMSE} = \EE \left[ \tilde{\sv}^{(l)}_k \tilde{\sv}^{(l)^\herm}_k \right]^{-1} \sqrt{q_{kl}}\sqrt{p_{kl}} M \Dm_k \betav^{(l)}_{kl},
$
\item The $l^{\rm th}$ BS computes the LSFD estimates $ \hat{s}_{kl,{dec}}=\av_{kl,{dec}}^\herm \tilde{\sv}^{(l)}_k,k=1,K$.
\item  The $l^{\rm th}$ BS estimates ${\rm SINR}_{kl},k=1,K$, (over multiple realizations)  and sends them to the corresponding users.
\item The $k^{\rm th}$ user in the $l^{\rm th}$ cell computes its new transmit power as
\begin{equation} \label{eqn:pwr-rule}
q_{kl}^{(n)} = \left\{ \begin{array}{cc}
q_{kl}^{(n-1)} \frac{\gamma}{{\rm SINR}_{kl}^{(n-1)}}, & \frac{q_{kl}^{(n-1)}}{{\rm SINR}_{kl}^{(n-1)}} \leq \frac{Q_{\max}}{\gamma}\\
\frac{Q_{\max}^2}{\gamma} \frac{{\rm
SINR}_{kl}^{(n-1)}}{q_{kl}^{(n-1)}}, & \frac{q_{kl}^{(n-1)}}{{\rm
SINR}_{kl}^{(n-1)}} > \frac{Q_{\max}}{\gamma}
\end{array} \right.
\end{equation}
\item\label{step:loop_ends} Assign $n=n+1$;
\end{enumerate}
\noindent{\bf The end.}

\begin{thm}\label{thm:power optimization converg}
The algorithm always converges and when (\ref{eqn:optz-q-3-limited}) is feasible, it converges to the optimal powers $q_{kl}$.
\end{thm}

In order to prove this theorem, we notice that (\ref{eqn:pwr-rule}) resembles the power update function of \cite{rasti2011distributed}, Definition 1. To prove the convergence of the decentralized algorithm, we first show that $\Id_{kl}(\qv) = \frac{q_{kl}}{{\rm SINR}_{kl}}$ and $\frac{1}{\Id_{kl}(\qv)} = \frac{{\rm SINR}_{kl}}{q_{kl}}$ are two-sided scalable functions and then invoke the result from \cite{rasti2011distributed}, Theorem 1, to complete the proof. The full proof can be found in Appendix \ref{appendix-two-sided-scalability}.

\section{Numerical Results} \label{sec:results}

 We consider a network consisting of $L = 19$ cells of
radius $1$ km wrapped into a torus (see \cite{NetworkMIMO3}). The wrapping allows us to
mimic an infinite network of cells. We assume that
   $K = 5$ and $M = 100$. For decentralized LSFD, we set
$L' = 6$. The maximum transmit power of each user during the pilot
and data transmission phase is set to $P_{\max} = Q_{\max} = 200$ mW, and the
large scale fading coefficients $\beta_{jkl}$ are computed according to
(\ref{eqn:pathloss}), with $\sigma_{\rm shad}^2 = 8$ dB.

\begin{figure}[!h]
\vspace{-2cm}
  \centering
\includegraphics[scale=0.4]{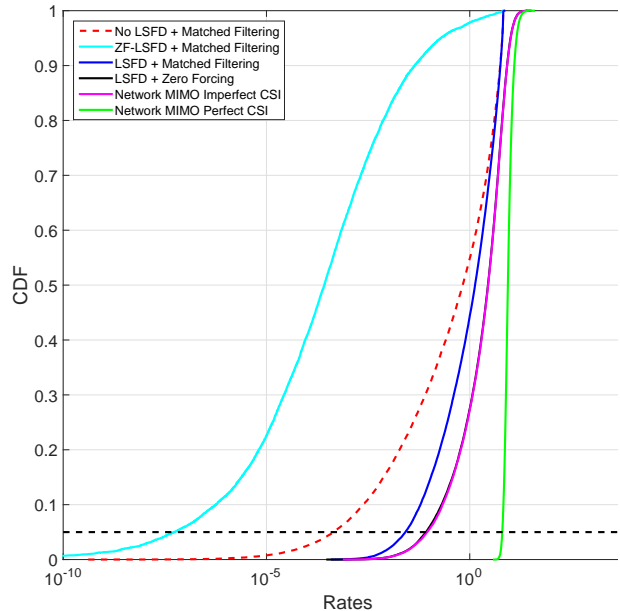}
\vspace{-1cm}
  \caption{CDF of the user rates for various schemes. Note that LSFD + Zero Forcing coincides with Network MIMO Imperfect
CSI, and ZF-LSFD with Matched Filtering shows very poor performance.}\label{fig1}
\end{figure}

\begin{figure}[!h]
\vspace{-3cm}
  \centering
\includegraphics[scale=0.7]{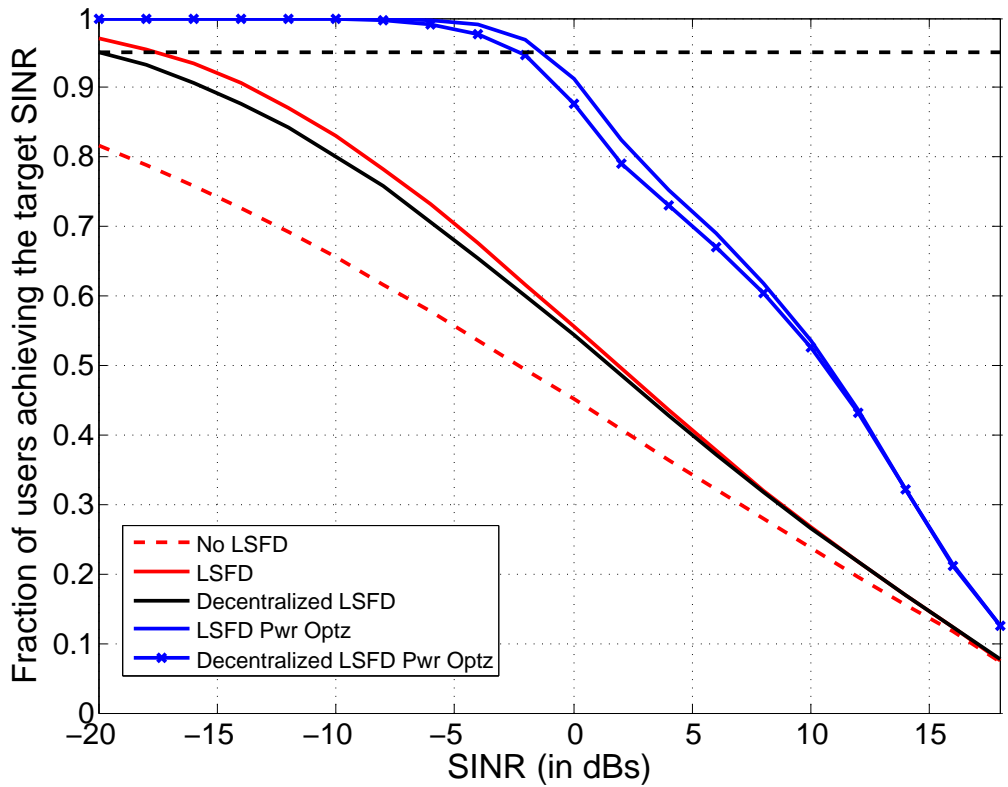}
\vspace{-4.5cm}
  \caption{Power optimization with varying target SINR}\label{fig2}
  \vspace{-0.8cm}
\end{figure}

Figure \ref{fig1} shows the CDF of the achievable rates  for the
various schemes considered in the paper. We mark the $5$ \% outage
rates by the dashed ``black'' horizontal line.
In addition to the results presented in Theorems \ref{thm:optimalSINR for MF} and \ref{thm:optimalSINR for ZF}, we also
 derived SINR expressions for a Network MIMO scheme where the BSs cooperate by
sharing between themselves all the channel state information. We
considered two variants of the scheme, one where the BSs  have access (magically) to the
actual $\gv_{lkj}$ (Network MIMO Perfect CSI) and the other where they only have $\hat{\gv}_{lkj}$
defined in (\ref{eq:hhat_lkl}) (Network MIMO Imperfect CSI). The analysis of these results is omitted due to space limitations, but we
use them in Fig. \ref{fig1}.

We observe that ZF-LSFD defined in (\ref{eq:ZF-LSFD}) performs very poorly even in comparison with no LSFD case. ZF-LSFD starts visibly gain only at $M>10^6$. At the same time we observe a $62.5$
fold increase in the $5$ \% outage rates when going from no LSFD
(``dashed red'') to LSFD (``blue'') with matched filtering and transmit power optimization.  When the
BS uses LSFD (``black'') with zero forcing and transmit power optimization, a $140$
fold increase is observed, showing that the obtained gains are truly
significant. It is also remarkable to see that LSFD with zero forcing
performs close to full cooperation with imperfect CSI.

Figure  \ref{fig2} shows the fraction of users achieving a certain
target SINR for varying target SINRs for global and decentralized LSFDs (with matched filtering) with and without power optimization. Again,
by looking at the $5$ \% outage rates, we observe a $16$ dB provided by the
transmit power optimization (
``blue'' curves). We observe only a minor
$0.5$ dB loss in going from global LSFD to decentralized LSFD, as for the case of power optimization as well as without it.

\section{Conclusion}
Large Scale Fading Decoding allows one to overcome the {\em pilot contamination} effect.
LSFD assumes a two level structure. First, BSs locally (independent from each other) conduct $M$-dimensional linear decoding and obtain
first level estimates of the transmitted uplink signals. Next, a network controller collects these estimates and conducts a second level linear decoding, which is based solely on the large scale fading coefficients between BSs and users.

In this paper, we considered  LSFDs with two $M$-dimensional linear decodings: matched filtering and zero-forcing. We first derived SINR expressions as functions of an LSFD decoding matrix used by the network controller. We further derive optimal LSFD decoding matrices that maximize SINRs of all users simultaneously. Next, we proposed a decentralized version of LSFD in which
 only a small number of neighboring base stations participate in LSFD. The problem of finding optimal LSFD matrices is  significantly more difficult in this case. One of the reasons for this is that transmit powers of users located outside of the neighboring cells is not known. We found a way around this problem and proposed a technique for empirical computation of optimal matrices for decentralized LSFD. Finally, we proposed a decentralized algorithm for uplink transmit power optimization, which provides additional system performance gain.


  Simulation results show that LSFD with zero forcing $M$-dimensional decoding and power optimization gives a 140-fold gain over MIMO systems without LSFD.

\section{Appendix}\label{sec:append}

\subsection{Proof of Theorem \ref{thm:SINR for MF}} \label{appendix-thm-2}
We start by computing the power of useful signal:
\begin{align*}
\EE\left[ \left|{\rm Useful \ \ Signal}\right|^2   \right]=\ & \EE \left[ |s_{kl}|^2 \right] q_{kl} \left|\sum_{j=1}^L a_{klj}^* \left( \EE [\hat{\gv}_{jkj}^\herm \hat{\gv}_{jkl} ] + \EE [\hat{\gv}_{jkj}^\herm \ev_{jkl} ]\right)\right|^2\nonumber\\
\stackrel{(a)}{=} \ &q_{kl} \left|\sum_{j=1}^L a_{klj}^* \frac{\beta_{jkl} \sqrt{p_{kl}}}{\beta_{jkj} \sqrt{p_{kj}}} \EE[\hat{\gv}_{jkj}^\herm \hat{\gv}_{jkj}]\right|^2\stackrel{(b)}{=} q_{kl} \left|\sum_{j=1}^L a_{klj}^* \frac{\beta_{jkj} \sqrt{p_{kj}} \beta_{jkl} \sqrt{p_{kl}} }{1 + \sum_{i=1}^L \beta_{jki} p_{ki}} M\right|^2,
\end{align*}
where $(a)$ follows from the fact that $\EE[\hat{\gv}_{jkj}^\herm \ev_{jkj}] = 0$ and (\ref{eqn:mmse-1}), and $(b)$ follows from
(\ref{eq:C_hv}).
Next
\begin{align}
&\mathrm{Var}\left[ {\rm Pilot \ \ Contamination}   \right]= \sum_{n=1,n\neq l}^L \left|\sum_{j=1}^L a_{klj}^* \left( \EE[\hat{\gv}_{jkj}^\herm \hat{\gv}_{jkn}] + \EE[\hat{\gv}_{jkj}^\herm \ev_{jkn}] \right) \sqrt{q_{kn}}\right|^2 \EE [| s_{kn}|^2]\nonumber\\
=\ &\sum_{n=1,n\neq l}^L \left|\sum_{j=1}^L a_{klj}^* \frac{\beta_{jkn} \sqrt{p_{kn}}}{\beta_{jkj} \sqrt{p_{kj}}} \EE[\hat{\gv}_{jkj}^\herm \hat{\gv}_{jkj}] \right|^2 q_{kn} = \sum_{n=1,n\neq l}^L \left|\sum_{j=1}^L a_{klj}^* \frac{\beta_{jkn} \sqrt{p_{kn}}}{\beta_{jkj} \sqrt{p_{kj}}} \frac{\beta_{jkj}^2 p_{kj}}{1 + \sum_{i=1}^L \beta_{jki} p_{ki}} M \right|^2 q_{kn} \nonumber\\
=\ &\sum_{n=1,n\neq l}^L \left|\sum_{j=1}^L a_{klj}^* \frac{\beta_{jkn} \sqrt{p_{kn}} \beta_{jkj} \sqrt{p_{kj}} }{1 + \sum_{i=1}^L \beta_{jki} p_{ki}} M \right|^2 q_{kn} \label{eq:pilot cont}
\end{align}
Finally,
\begin{align}
&\mathrm{Var}\left[ {\rm Interference \ \ plus \ \ Noise \ \ Terms}   \right] \nonumber\\
= &\sum_{n=1}^L \EE \left[ \left| \sum_{j=1}^L a_{klj}^* \left( \hat{\gv}_{jkj}^\herm \hat{\gv}_{jkn} - \EE[\hat{\gv}_{jkj}^\herm \hat{\gv}_{jkn}] \right) \right|^2 \right] q_{kn} \EE [ |s_{kn}|^2]  \nonumber\\
&+ \sum_{n=1}^L \sum_{m \neq k} \EE \left[ \left| \sum_{j=1}^L a_{klj}^* \hat{\gv}_{jkj}^\herm \gv_{jmn} \right|^2 \right] q_{mn} \EE [ |s_{mn}|^2 ] + \EE \left[ \left| \sum_{j=1}^L a_{klj}^* \hat{\gv}_{jkj}^\herm \zv_{j} \right|^2 \right]\nonumber\\
= &\sum_{n=1}^L \sum_{j=1}^L |a_{klj}|^2 \frac{\beta_{jkn}^2 p_{kn}}{\beta_{jkj}^2 p_{kj}}
\EE\left[ \left| \hat{\gv}_{jkj}^\herm \gv_{jkj} - \EE[\hat{\gv}_{jkj}^\herm \gv_{jkj}]\right| ^2\right]q_{kn} + \sum_{n=1}^L \EE \left[ \left| \sum_{j=1}^L a_{klj}^* \hat{\gv}_{jkj}^\herm \ev_{jkn} \right|^2 \right] q_{kn} \nonumber\\
& + \sum_{n=1}^L \sum_{m \neq k} \EE \left[ \left| \sum_{j=1}^L a_{klj}^* \hat{\gv}_{jkj}^\herm \gv_{jmn} \right|^2 \right] q_{mn} + \EE \left[ \left| \sum_{j=1}^L a_{klj}^* \hat{\gv}_{jkj}^\herm \zv_{j} \right|^2 \right]
 \label{eq:Interf Plus Noise}
\end{align}
To compute the first term in this expression, we note that, using (\ref{eq:C_hv}), we have
\begin{align*}
\EE \left| [ (\hat{\gv}_{jkj}^\herm \hat{\gv}_{jkn})] \right|^2&=\left( \frac{\beta_{jkj}^2 p_{kj}}{1 + \sum_{i=1}^L \beta_{jki} p_{ki}} \right)^2(M+M^2),~
\EE[\hat{\gv}_{jkj}^\herm \gv_{jkj}]= \frac{\beta_{jkj}^2 p_{kj}}{1 + \sum_{i=1}^L \beta_{jki} p_{ki}}M
\end{align*}
giving
$
\mathrm{Var} [ (\hat{\gv}_{jkj}^\herm \hat{\gv}_{jkn}) ] = \left( \frac{\beta_{jkj}^2 p_{kj}}{1 + \sum_{i=1}^L \beta_{jki} p_{ki}} \right)^2 M \nonumber
$.
Hence the first term in (\ref{eq:Interf Plus Noise}) is equal to
$$
\sum_{n=1}^L\sum_{j=1}^L |a_{klj}|^2 \frac{\beta_{jkn}^2 p_{kn}}{\beta_{jkj}^2 p_{kj}} \left( \frac{\beta_{jkj}^2 p_{kj}}{1 + \sum_{i=1}^L \beta_{jki} p_{ki}}\right)^2M
$$
We compute other terms in (\ref{eq:Interf Plus Noise}) in a similar way, and, after some calculations, obtain
\begin{align*}
& \mathrm{Var}\left[ {\rm Inter. \ \ plus \ \ Noise \ \ Terms}  \right]
= \sum_{j=1}^L |a_{klj}|^2 \frac{\beta_{jkj}^2 p_{kj}}{1 + \sum_{i=1}^L \beta_{jki} p_{ki}} M \left( \sum_{n=1}^L \sum_{m = 1}^K  \beta_{jmn} q_{mn} + 1 \right)
\end{align*}
Combining these expressions and using (\ref{eq:hat_a}), we obtain the claim.\qed

\subsection{Proof of Theorem \ref{thm:power optimization converg}} \label{appendix-two-sided-scalability}

We provide here a proof of the two sided scalability of the functions $\Id_{kl}(\qv) = \frac{q_{kl}}{{\rm SINR}_{kl}}$ and $\frac{1}{\Id_{kl}(\qv)} = \frac{{\rm SINR}_{kl}}{q_{kl}}$. A function $f(\xv)$ is a two-sided scalable function \cite{rasti2011distributed} if it satisfies the
property: {\em for all $\alpha > 1$ and vectors $\xv_1,\xv_2$, $\frac{1}{\alpha} \xv_1 \leq \xv_2 \leq \alpha \xv_1$ implies
$
\frac{1}{\alpha} f(\xv_1) < f(\xv_2) < \alpha f(\xv_1)
$.}

In order to prove that $\Id_{kl}(\qv)$ and $\frac{1}{\Id_{kl}(\qv)}$ are two-sided scalable, we first need to show that $\Id_{kl}(\qv)$ is a standard interference function \cite{yates1995framework}, which means that it satisfies the following three properties:
1) $\Id_{kl}(\qv) \geq 0\ \ \forall\ \ \qv \geq \zerov$, 2) $\Id_{kl}(\qv_1) \geq \Id_{kl}(\qv_2)$, 3)
for any $\alpha > 1$, $\Id_{kl}(\alpha \qv) < \alpha \Id_{kl}(\qv)$.

Clearly, $\Id_{kl}(\qv) \geq 0$ since both $q_{kl}$ and ${\rm SINR}_{kl}$ are positive quantities.
Using (\ref{eq:Ritz_theorem}), we obtain
\begin{align}\label{eq:q1>q2}
\Id_{kl}(\qv) = \frac{q_{kl}}{{\rm SINR}_{kl}}= &\frac{1}{\betav_{kl}^{' \herm} \left( \sum_{n=1,n \neq l}^L \betav'_{kn} \betav_{kn}^{' \herm} p_{kn} q_{kn} M + \Lambdam'_k (\qv) \right)^{-1} \betav'_{kl} p_{kl} M} \nonumber\\
=& \left[ \max_{\uv} \frac{\uv^\herm \betav'_{kl} \betav_{kl}^{' \herm} \uv p_{kl} M }{\uv^\herm \left( \sum_{n=1,n\neq l}^L \betav'_{kn} \betav_{kn}^{' \herm} p_{kn} q_{kn} M + \Lambdam'_k (\qv) \right) \uv} \right]^{-1}.
\end{align}

If $\qv_1 \geq \qv_2$, then from (\ref{eq:lambda}), it follows that $\Lambda_k'(\qv_1)-\Lambda_k'(\qv_2)$ is positive definite. Denote by $q_{1,kn}$ and $q_{2,kn}$ the corresponding entries of $\qv_1$ and $\qv_2$. The matrices $\betav'_{kn} \betav_{kn}^{' \herm}p_{kn}(q_{1,kn}-q_{2,kn})$
are also positive definite. Hence, we have for any vector $\uv$:
\begin{align*}
\uv^\herm \left( \sum_{n=1,n\neq l}^L \betav'_{kn} \betav_{kn}^{' \herm} p_{kn} q_{1,kn} M + \Lambdam'_{k} (\qv_1) \right) \uv
 \geq \uv^\herm \left( \sum_{n=1,n\neq l}^L \betav'_{kn} \betav_{kn}^{' \herm} p_{kn} q_{2,kn} M + \Lambdam'_{k} (\qv_2) \right) \uv
\end{align*}
Let
$$
\tilde{\uv} = {\rm arg} \max_{\uv} \frac{\uv^\herm \betav'_{kl} \betav_{kl}^{' \herm} \uv p_{kl} M }{\uv^\herm \left( \sum_{n=1,n\neq l}^L \betav'_{kn} \betav_{kn}^{' \herm} p_{kn} q_{1,kn} M + \Lambdam'_k (\qv_1) \right) \uv}
$$
Now, from (\ref{eq:q1>q2}), it follows that
\begin{align*}
&\max_{\uv} \frac{\uv^\herm \betav'_{kl} \betav_{kl}^{' \herm} \uv p_{kl} M }{\uv^\herm \left( \sum_{n=1,n\neq l}^L \betav'_{kn} \betav_{kn}^{' \herm} p_{kn} q_{1,kn} M + \Lambdam'_k (\qv_1) \right) \uv} \leq \frac{\tilde{\uv}^\herm \betav'_{kl} \betav_{kl}^{' \herm} \tilde{\uv} p_{kl} M }{\tilde{\uv}^\herm \left( \sum_{n=1,n\neq l}^L \betav'_{kn} \betav_{kn}^{' \herm} p_{kn} q_{1,kn} M + \Lambdam'_k (\qv_1) \right) \tilde{\uv}} \nonumber\\
\leq& \frac{\tilde{\uv}^\herm \betav'_{kl} \betav_{kl}^{' \herm} \tilde{\uv} p_{kl} M }{\tilde{\uv}^\herm \left( \sum_{n=1,n\neq l}^L \betav'_{kn} \betav_{kn}^{' \herm} p_{kn} q_{2,kn} M + \Lambdam'_k (\qv_2) \right) \tilde{\uv}}\leq \max_{\uv} \frac{\uv^\herm \betav'_{kl} \betav_{kl}^{' \herm} \uv p_{kl} M }{\uv^\herm \left( \sum_{n=1,n\neq l}^L \betav'_{kn} \betav_{kn}^{' \herm} p_{kn} q_{2,kn} M + \Lambdam'_k (\qv_2) \right) \uv}
\end{align*}
From this, it follows that
$
\Id_{kl}(\qv_1) \geq \Id_{kl}(\qv_2).
$
Finally, for any $\alpha > 1, \uv$ and $j$, we have
\begin{align*}
\alpha (1 + \sum_{i=1}^L \beta_{jki} p_{ki}) \left( 1 + \sum_{n=1}^L \sum_{m = 1}^K \beta_{jmn} q_{mn} \right)> (1 + \sum_{i=1}^L \beta_{jki} p_{ki}) \left( 1 + \sum_{n=1}^L \sum_{m = 1}^K \beta_{jmn} \alpha q_{mn} \right)
\end{align*}
This implies $\alpha \uv^\herm \Lambdam'_{k}(\qv) \uv > \uv^\herm \Lambdam'_{k}(\alpha \qv)$ and further
\begin{align*}
\alpha \uv^\herm \left( \sum_{n=1,n\neq l}^L \betav'_{kn} \betav_{kn}^{' \herm} p_{kn} q_{kn} M + \Lambdam'_{k} (\qv_1) \right) \uv > \uv^\herm \left( \sum_{n=1,n\neq l}^L \betav'_{kn} \betav_{kn}^{' \herm} p_{kn} \alpha q_{kn} M + \Lambdam'_{k} (\alpha \qv) \right) \uv
\end{align*}
Hence,
\begin{align*}
&\max_{\uv} \frac{\uv^\herm \betav'_{kl} \betav_{kl}^{' \herm} \uv p_{kl} M }{\alpha \uv^\herm \left( \sum_{n=1,n\neq l}^L \betav'_{kn} \betav_{kn}^{' \herm} p_{kn} q_{kn} M + \Lambdam'_k (\qv) \right) \uv}
\leq \frac{\tilde{\uv}^\herm \betav'_{kl} \betav_{kl}^{' \herm} \tilde{\uv} p_{kl} M }{\alpha \tilde{\uv}^\herm \left( \sum_{n=1,n\neq l}^L \betav'_{kn} \betav_{kn}^{' \herm} p_{kn} q_{kn} M + \Lambdam'_k (\qv) \right) \tilde{\uv}} \nonumber\\
<& \frac{\tilde{\uv}^\herm \betav'_{kl} \betav_{kl}^{' \herm} \tilde{\uv} p_{kl} M }{ \tilde{\uv}^\herm \left( \sum_{n=1,n\neq l}^L \betav'_{kn} \betav_{kn}^{' \herm} p_{kn} \alpha q_{kn} M + \Lambdam'_k (\alpha \qv) \right) \tilde{\uv}}
\leq \max_{\uv} \frac{\uv^\herm \betav'_{kl} \betav_{kl}^{' \herm} \uv p_{kl} M }{\uv^\herm \left( \sum_{n=1,n\neq l}^L \betav'_{kn} \betav_{kn}^{' \herm} p_{kn} \alpha q_{kn} M +
\Lambdam'_k (\alpha \qv) \right) \uv}
\end{align*}
From this, it follows that
$
\alpha \Id_{kl}(\qv) > \Id_{kl}(\alpha \qv).
$
Thus, we proved that $\Id_{kl}(\qv)$ is a standard interference function.
Replacing $\qv$ by $\frac{1}{\alpha} \qv$ in Property 3 of the standard interference function, we have
\begin{eqnarray}
\Id_{kl}(\qv) < \alpha \Id_{kl}\left(\frac{1}{\alpha} \qv \right) &\Longrightarrow& \frac{1}{\alpha} \Id_{kl}(\qv) < \Id_{kl}\left(\frac{1}{\alpha} \qv\right) \label{eqn:prop-3-b}
\end{eqnarray}
Now, for any $\qv_1$ and $\qv_2$ and all $\alpha > 1$ such that $\frac{1}{\alpha} \qv_1 \leq \qv_2 \leq \alpha \qv_1$, we have
$$
 \frac{1}{\alpha} \Id_{kl}(\qv_1) \stackrel{(a)}{<} \Id_{kl}\left(\frac{1}{\alpha}\qv_1\right) \stackrel{(b)}{\leq} \Id_{kl}(\qv_2) \stackrel{(c)}{\leq} \Id_{kl}(\alpha \qv_1) \stackrel{(d)}{<} \alpha \Id_{kl}(\qv_1)
\Longrightarrow \frac{1}{\alpha} \Id_{kl}(\qv_1) < \Id_{kl}(\qv_2) < \alpha \Id_{kl}(\qv_1)
$$
where $(a)$ follows from (\ref{eqn:prop-3-b}), $(b)$ and $(c)$ follow from Property 2 of the standard interference function, and $(d)$ follows from Property 3. Thus, $\Id_{kl} (\qv)$ is a standard interference function.
In a similar fashion, we have
$$
 \frac{1}{\alpha} \frac{1}{\Id_{kl}(\qv_1)} \stackrel{(a)}{>} \frac{1}{\Id_{kl}(\alpha \qv_1)} \stackrel{(b)}{\geq} \frac{1}{\Id_{kl}(\qv_2)} \stackrel{(c)}{\geq} \frac{1}{\Id_{kl}\left(\frac{1}{\alpha} \qv_1\right)} \stackrel{(d)}{>} \alpha \frac{1}{\Id_{kl}(\qv_1)}
\Longrightarrow \frac{1}{\alpha} \frac{1}{\Id_{kl}(\qv_1)} < \frac{1}{\Id_{kl}(\qv_2)} < \alpha \frac{1}{\Id_{kl}(\qv_1)}
$$
where $(a)$ follows from Property 3 of the standard interference function, $(b)$ and $(c)$ follow from Property 2, and $(d)$ follows from (\ref{eqn:prop-3-b}). Thus, $\frac{1}{\Id_{kl} (\qv)}$ is also a standard interference function.


\end{document}